\documentclass[]{aa}

\usepackage[]{rotating}
\usepackage[]{graphicx}

\input epsf        % defines \epsfbox and supporting macros

\begin{document}
%\refereelayout

\title{Modelling the spectral response of the {\it \bfseries Swift}-XRT
  CCD camera: Experience learnt from in-flight calibration}

\author{O. Godet$^1$, A. P. Beardmore$^1$, A. F. Abbey$^1$, J. P. Osborne$^1$,
G. Cusumano$^2$, C. Pagani$^3$, M. Capalbi$^4$, M. Perri$^4$, K. L. Page$^1$,
D. N. Burrows$^3$, S. Campana$^5$, J. E. Hill$^{6,7}$, J. A. Kennea$^3$, A. Moretti$^5$}

\offprints{og19@star.le.ac.uk}

\institute{$^1$ X-ray and Observational Astronomy Group, Department of
Physics \& Astronomy, University of Leicester, LE1 7RH, UK\\
$^2$ INAF-Istituto di Astrofisica Spaziale e Fisica Cosmica Sezione di
Palermo, Via U. La Malfa 153, 90146 Palermo, Italy \\
$^3$ Department of Astronomy \& Astrophysics, 525 Davey Lab, Pennsylvania
State University, University Park, PA 16802, USA\\
$^4$ ASI Science Data Center, Via G. Galilei, I-00044 Frascati, Italy \\
$^5$ INAF-Osservatorio Astronomico di Brera, Via E. Bianchi 46, 23807, Merate (LC), Italy \\
$^6$ CRESST Goddard Space Flight Center, Greenbelt, Maryland 20771, USA \\
$^7$ Universities Space Research Association, 10211 Wincopin Circle, Suite
500, Columbia, MD, 21044-3432, USA\\}

\date{Accepted : 2008 November 21}

\titlerunning{Modelling of the {\it \bfseries Swift}-XRT spectral response }
\authorrunning{Godet et al.}

\abstract{ {\it Context.}  Since its launch in November 2004, {\it Swift} has
revolutionised our understanding of gamma-ray bursts.  The X-ray telescope
(XRT), one of the three instruments on board {\it Swift}, has played a key
role in providing essential positions, timing, and spectroscopy of more than
300 GRB afterglows to date. Although {\it Swift} was designed to observe GRB
afterglows with power-law spectra, {\it Swift} is spending an increasing
fraction of its time observing more traditional X-ray sources, which have more
complex spectra.

{\it Aims.}  The aim of this paper is a detailed description of the CCD
response model used to compute the XRT RMFs (redistribution matrix files), the
changes implemented to it based on measurements of celestial and on-board
calibration sources, and current caveats in the RMFs for the spectral analysis
of XRT data.

{\it Results.}  We show that the XRT spectral response calibration was
complicated by various energy offsets in photon counting (PC) and windowed
timing (WT) modes related to the way the CCD is operated in orbit (variation
in temperature during observations, contamination by optical light from the
sunlit Earth and increase in charge transfer inefficiency).  We describe how
these effects can be corrected for in the ground processing software. We show
that the low-energy response, the redistribution in spectra of absorbed
sources, and the modelling of the line profile have been significantly
improved since launch by introducing empirical corrections in our code when it
was not possible to use a physical description.  We note that the increase in
CTI became noticeable in June 2006 (\emph{i.e.} 14 months after launch), but
the evidence of a more serious degradation in spectroscopic performance (line
broadening and change in the low-energy response) due to large charge traps
(\emph{i.e.} faults in the Si crystal) became more significant after March
2007. We describe efforts to handle such changes in the spectral
response. Finally, we show that the commanded increase in the substrate
voltage from 0 to 6\,V on 2007 August 30 reduced the dark current, enabling
the collection of useful science data at higher CCD temperature (up to
-50$^\circ$C). We also briefly describe the plan to recalibrate the XRT
response files at this new voltage.

{\it Conclusions.}  We show that the XRT spectral response is described well
by the public response files for line and continuum spectra in the 0.3-10 keV
band in both PC and WT modes.

}

\maketitle

\section{Introduction}

Successfully launched on 2004 November 20, the {\it Swift} gamma-ray burst
satellite (Gehrels et al. 2004) consists of three instruments: the wide-field
of view, gamma-ray burst alert telescope (BAT; Barthelmy et al. 2005) and two
narrow field instruments (NFIs), the X-ray telescope (XRT; Burrows et
al. 2005) and the UV/optical telescope (UVOT; Roming et al. 2005). Thanks to
the unique ability of {\it Swift} to slew automatically after the BAT trigger,
bursts are typically within the field of view of the narrow field instruments
within a couple of minutes after the trigger. Thus, the XRT routinely provides
positions with an accuracy of a few arc-seconds to the GRB community
world-wide and measures the early X-ray light-curves and spectra of most GRB
afterglows at which it is promptly pointed. Up to 2008 August 11, a total of
357 GRBs were detected by the BAT, of which 314 GRBs were observed by the XRT,
although only 297 X-ray afterglows were thus detected by the XRT. Of these
burst observations, 253 were prompt slews (i.e. less than 300\,s after the BAT
trigger), and the XRT detected 94\%\,\footnote{see
http://swift.gsfc.nasa.gov/docs/swift/archive/grb$_{-}$table/
grb$_{-}$stats.php} of those bursts (i.e. 238 afterglows).

The {\it Swift}-XRT spectro-temporal observations associated with on-board and
ground-based multi-wavelength observations have actively driven GRB science in
the past three years by shedding new light on the physics of these
objects. X-ray observations have revealed previously unexpected behaviour,
including: multiple temporal breaks observed in some GRB light-curves
inconsistent with the standard afterglow models (e.g. Zhang et al. 2006,
Nousek et al. 2006, O'Brien et al. 2006, Willingale et al. 2007) and the
discovery of X-ray flares observed in about $50\%$ of the {\it Swift}
afterglows (e.g. Falcone et al. 2006, 2007; Chincarini et al. 2007; Goad et
al. 2007; Godet et al. 2007a, 2006; Burrows et al. 2007). Recent work has
shown that some spectra cannot be fit by simple absorbed power-laws due to
curvature, which can be interpreted as spectral breaks caused by the temporal
shift of the energy peak through the XRT energy band during X-ray flares
(e.g. GRB 051117A: Goad et al. 2007 and GRB 050822: Godet et al. 2007a). In
some other cases, the presence of an extra component was suggested (see Butler
2007, Moretti et al. 2008a for a general study) such as a blackbody component,
which could be interpreted as: possible emission of a jet cocoon; the
first-ever detection of a shock breakout from a massive star, likely a
Wolf-Rayet, in the peculiar event GRB 060218 (Campana et al. 2006); possible
photospheric emission for some X-ray flares in GRB 050822 (Godet et
al. 2007a).

More detailed X-ray spectral analyses have shed light on the environment of
GRBs and the nature of the progenitor itself. Campana et al. (2007)
showed evidence that the progenitor of the high redshift GRB 050904 at $z=6.3$
was located in a dense molecular cloud with a metallicity $Z >
0.03~Z_\odot$. Using detailed modelling of the low-energy part of the X-ray
spectrum, Campana et al. (2008a) showed that the progenitor of GRB 060218 was
likely to be a massive star characterized by a fast stellar rotation and
initial sub-solar metallicity giving, for the first time, direct evidence
about the properties of GRB progenitors.

Thanks to its large energy band-pass and the ability to rapidly schedule
targets, {\it Swift} is also a powerful tool for obtaining essential insights
into the physics of non-GRB objects, especially transients. For example, the
XRT has provided essential spectroscopic information since the start of the
outburst of the recurrent nova RS Ophiuchi in 2006, enabling, for the first
time, detailed observations of the evolution of the Super-Soft Component in
such an object (e.g. Bode et al. 2006, Hachisu et al. 2007). It also provided
important spectroscopic information during the follow-up of several transient
events (e.g., blazars: Tramacere et al. 2007; X-ray binaries: Esposito et
al. 2007, Wijnands et al. 2007, Rykoff et al. 2007, Romano et al. 2007,
Brocksopp et al. 2005; Comet 9P/Tempel 1: Willingale et al. 2006; active
galactic nuclei: Molina et al. 2007, Tueller et al. 2007). From the beginning
of 2008, the fraction of time spent on non-GRB science targets has been
$\sim 45\%$, and this will increase in the future.

\begin{figure}[h]
\begin{center}
\includegraphics[height=5.5cm]{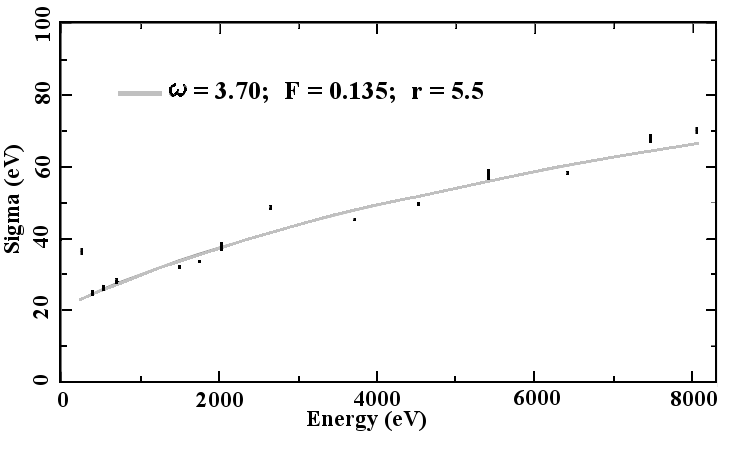}
\end{center}
\caption[]{Energy resolution ($\sigma=\frac{\rm FWHM}{2.35}$) versus energy as
  measured during the ground calibration at the Leicester calibration facility
  using the in-flight CCD-22 cooled at -100$^\circ$C. The resolution was
  measured over the central $200\times 200$ pixel window. The curve
  corresponds to the theoretical prediction given by $\sigma = \omega
  \sqrt{\frac{FE}{\omega}+r^2}$ where $F$, $E$, $\omega$ and $r$ are the Fano
  factor, the photon energy, the pair creation energy coefficient and the
  readout electronic noise, respectively.  The thermal noise could be
  neglected with respect to the readout electronic noise due to the low
  temperature used during the measurements.}
\label{fig_F}
\end{figure}

\begin{figure*}
\begin{center}
\begin{tabular}{c}
\includegraphics[height=14cm]{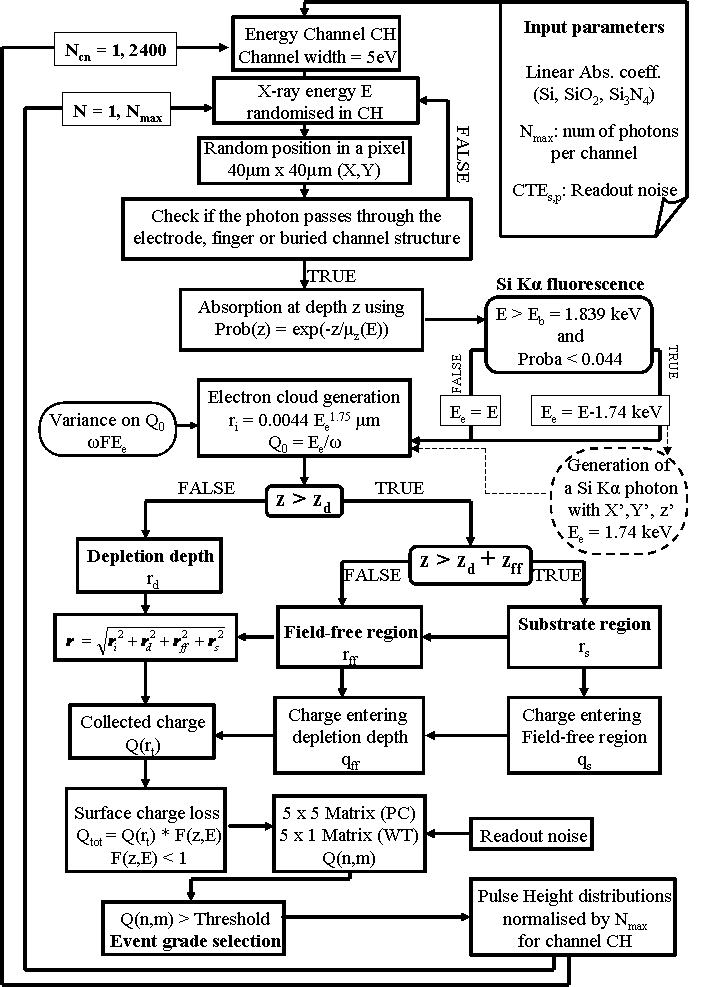}\\
\end{tabular}
\end{center}
\caption[]{Architecture of the CCD response model used to compute the XRT
  RMFs. The RMFs contain 2400 input energy channels with a 5\,eV energy
  width. In each energy channel, $\rm N_{\rm max} > 5\times 10^5$ photons with
  an energy randomised in a given bin are considered in order to avoid adding
  numerical noise to the computation.  $\mu_z(E)$ corresponds to the linear
  absorption coefficient in the silicon, which is a function of the energy of
  the incoming photons ($E$), while $CTE_{\it s,p}$ corresponds to the charge
  transfer efficiency during the readout process in the serial and parallel
  direction, respectively (see Section~\ref{CTI}). $z$, $z_d$, $z_{\it ff}$
  correspond to the depth of interaction of the incoming photons ($z=0$
  corresponding to the surface of the Si active bulk), the depletion depth (an
  input parameter) and the field-free region depth defined as $z_{\it ff} =
  \Delta z_{\it CCD} - z_d - z_s$ (where $z_s = 200\,\mu{\rm m}$ is the
  thickness of the substrate and $\Delta z_{\it CCD} = 280\,\mu{\rm m}$ is the
  thickness of the CCD).  $F = 0.115$ and $\omega = 3.68$ eV per pair
  electron/hole are the Fano factor and the pair creation energy coefficient
  in the silicon, respectively, while $Q_0$ corresponds to the initial charge
  generated by the interaction of an incoming photons within the device.  The
  different equations giving the total charge spreading radius ($r_t$) in the
  different layers of the CCD and the collected charge $\rm Q(r_t)$ can be
  found in the Section~\ref{model}.  $f(E,z)$ corresponds to a function
  describing the charge losses at the surface of the CCD (see
  Section~\ref{phys}). It depends on energy and depth of interaction of the
  incoming photons. Mostly photons interacting in the opened electrode area
  will suffer from surface charge losses (see the text in Section2.1.3).  The
  dashed lines correspond to the case where a 1.74 keV K$\alpha$ fluorescence
  photon is emitted following the interaction of an incoming photon with an
  energy $E > E_b = 1.839$ keV where $E_b$ is the electron binding energy for
  the Si K shell. The probability of generating such a fluorescence
  photon is 4.4\%.}
\label{fig_scheme}
\end{figure*}

The results above show that X-ray spectroscopy is a powerful and essential
tool for obtaining physical insights for celestial objects. These results
depend on the intrinsic performance of the detector and the calibration of its
spectral response. The XRT uses a front-illuminated e2v CCD-22, offering good
spectroscopic performance with an energy resolution (FWHM) before launch of
135 eV at 5.9 keV (see also Fig.~\ref{fig_F}). The CCD-22 is a three phase
frame transfer device, which utilises high resistivity (typically
$4000\,\Omega$ cm) silicon and an open electrode structure, originally
designed for, and used in, the EPIC (European Photon Imaging Camera) MOS
cameras on-board {\it XMM-Newton}. The open electrode structure significantly
improves the quantum efficiency of the device at low energy. The CCD imaging
area consists of a $600\times 600$ pixel array with a pixel size of $40\times
40\,\mu\mathrm{m}^2$. Forty microns corresponds to 2.36 arc-seconds in the XRT
field of view (e.g. Short et al. 2002).  To mitigate the effects of pile-up,
the CCD can automatically switch between different readout modes depending on
the source brightness once the spacecraft is settled on the source (Hill et
al. 2004, 2005):
\begin{itemize}
\item[1)] Photo-diode (PD) mode at the highest count rates with a 0.14 ms time
  resolution and no spatial information;
\item[2)] Windowed Timing (WT) mode at moderate count rates, which uses a 200
column window covering the central 8 arcminutes of the XRT field of view and
provides 1-D spatial imaging information. In that mode, the columns are
clocked continuously to provide timing information in the trailed image along
each column with a 1.8 ms time resolution, at the expense of imaging
information in this dimension (pixels are binned by a factor of 10 along
columns);
\item[3)] at lower count rates, Photon Counting mode (PC), which provides 2-D
spatial imaging information, but with a 2.5 s time resolution.
\end{itemize}
The PD mode has not been used since the CCD was hit by a particle
(micro-meteoroid) on 2005 May 27 (Carpenter et al. 2006), due to the
apparition of several bad columns on the detector, so we will focus only on
the PC and WT modes in the remainder of the paper.

The XRT effective area (EA) is made up of the response of the mirrors, the
filter placed in front of the CCD detector to reduce optical and UV photon
loading and the CCD detector response. The mirror and filter responses are
included in the auxiliary response files (ARFs), while the CCD response is
included in the redistribution matrix files (RMFs).

To avoid any misinterpretation in the spectral analysis, it is essential to
have a good understanding of the instrument response and its operational
limitations.  The aim of this paper is to describe in detail our CCD response
model used to compute the RMFs, the changes that have been made to the RMFs
and ARFs since the launch (based on in-flight calibration) and the caveats to
be aware of in the spectral analysis of XRT data when using the current
(version 11) response files distributed in the CALDB release on 2008-06-25. An
overview of the other aspects of the XRT calibration can be found in Campana
et al. (2008b) and a detailed study of the XRT background can be found in
Moretti et al. (2008b). This paper is organised as follows:
\begin{itemize}
\item[{\bf \S 2 -}] We describe the physics implemented in our RMF code. We
also briefly describe the spectroscopic performance of the RMFs in both PC and
WT modes prior to launch using ground calibration data.

\item[{\bf \S 3 -}] We discuss the in-flight operation of the XRT and its
impact on the calibration of the XRT response, the calibration program, and in
detail post-launch changes made to the CCD spectral model (low-energy
response, line shoulder, shelf, RMF redistribution) and the ARFs based on
celestial target calibration.

\item[{\bf \S 4 -}] We present the spectroscopic performance of the XRT
for several celestial targets compared to observations with other X-ray
instruments, as well as caveats (line broadening due to the build-up of
charge traps on the CCD and changes in the in-orbit operation of the CCD
caused by raising the substrate voltage from 0\,V to 6\,V in 2007 August 30)
for the spectral analysis when using the current RMFs and ARFs.

\item[{\bf \S 5 -}] We present the main conclusions of the paper.

\end{itemize}

%################################
\section{Computation of the spectral response}
\label{code}

\begin{figure*}
\begin{center}
\hspace{2cm}\includegraphics[width=14.cm,height=12cm]{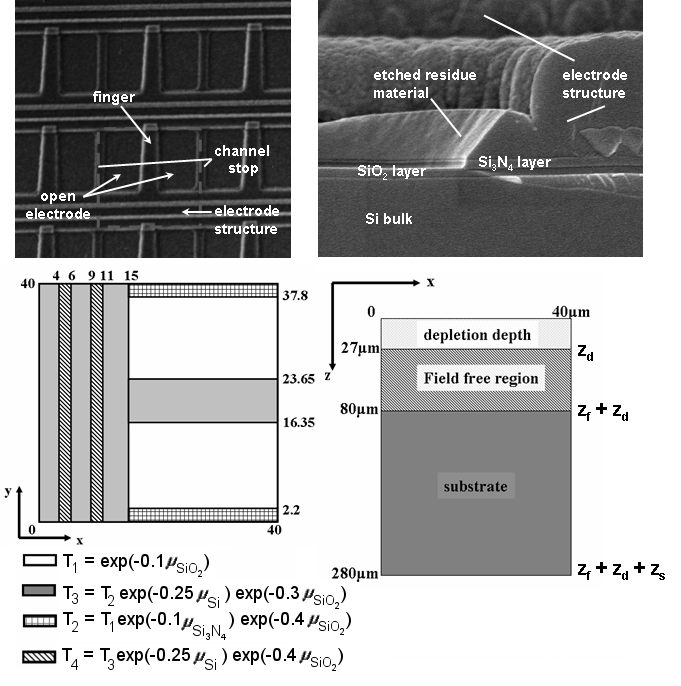}
\end{center}
\caption[]{Pictures of the CCD-22 structure and geometry of a CCD-22
  pixel as used in our spectral response code. (Top left and right panels)
  Top and side view of a CCD-22 pixel, respectively. These images were
  obtained by a scanning electron microscope provided by e2v. The different
  parts of the pixel are shown on these two images.  (Bottom left and right
  panels) Geometry of the CCD-22 pixel as used in our CCD response model (the
  dimensions are given in units of $\mu$m): (left) Top view of the simplified
  geometry of a CCD-22 pixel. The thicknesses of the compounds (Si, SiO$_2$
  and Si$_3$N$_4$) present in the finger and electrode structures are shown in
  the equations giving the transparency of each part of the electrode
  structure on top of the silicon bulk (the coefficients $T_{i=1,4}$); (right)
  Side view showing the different layers formed in the CCD (see text for more
  details). The epitaxial layer is $80\,\mu$m thick (p-type) and the substrate
  is $200\,\mu$m thick. The 0 in the $z$ coordinate corresponds to the top of
  the Si active bulk, the electrode being on top of it. The incident X-ray
  photons enter from the top.}
\label{fig_pixel}
\end{figure*}

The XRT spectral response is based on a physical model describing the
interaction of photons within the silicon bulk of a CCD-22 pixel via
Monte-Carlo simulations.  A Monte-Carlo simulator was developed at the
University of Leicester in order to generate the RMFs for all the XRT readout
modes (Short et al.  2002, Mukerjee et al. 2004).  The CCD response model was
refined by implementing ad-hoc corrections in our code based on the ground and
in-flight calibration. We discuss here the main steps used to generate the
RMFs.

\subsection{RMF generation}
\label{RMFgen}

\begin{figure}[h]
\begin{center}
\includegraphics[width=8.cm]{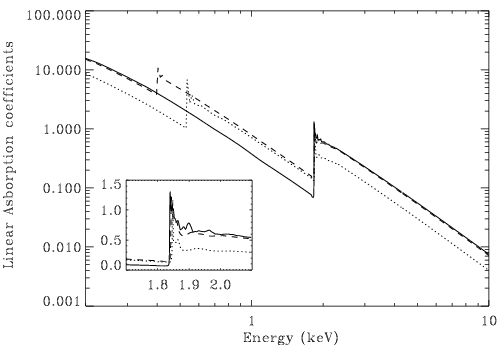}
\end{center}
\caption[]{Linear absorption coefficients of the CCD-22 compounds: (solid
  line) Si; (dotted line) SiO$_2$; (dashed line) Si$_3$N$_4$. The small inset
  shows the XAFS just above the Si edge. The energy of the Si edge is slightly
  different depending on the compound considered. This is due to differences
  in the crystal lattice structure.}
\label{fig_linear}
\end{figure}

Figure~\ref{fig_scheme} shows the architecture of our CCD response model.  To
compute the RMFs, we stack simulated spectra of monochromatic X-rays using
simple geometry of a CCD-22 pixel as input parameters, including the electrode
structure, the finger and the open area (see Fig.~\ref{fig_pixel}) and the
linear absorption coefficients of the different CCD compounds. The linear
absorption coefficients used in our simulations include the effects of the
rapid oscillatory X-ray Absorption Fine Structures (XAFS) just above the
absorption edges of the CCD constituents (mainly N, O and Si) as measured by
Owens et al. (1996a) for the JET-X CCD (see Fig~\ref{fig_linear}). It is
essential to take them into account to obtain proper modelling of the spectral
response around the edges.

\subsubsection{Carrier generation and charge spreading}
\label{model}

0.2-10 keV photons impinging on the detector have a high probability of
interacting via the photoelectric effect. This results in the generation of a
charge cloud with an initial charge $Q_0$ given by:
\begin{equation}
Q_0 = e\times \left(\frac{E_e}{\omega} + \sqrt{F~\omega~E_e}\times R\right)
\end{equation}
where $F$ and $\omega$ correspond to the Fano factor and the pair creation
energy coefficient in the silicon, respectively. $R$ is a random number
uniformly distributed between -1 and 1; $e$ is the elementary charge of
the electrons ($e = 1.6\times 10^{-19}$ C). If the incoming photons have an
energy $E$ above $E_b = 1.839$ keV (i.e. the binding energy of the K-shell
electrons in the silicon), then there is a 4.4\% probability of producing a Si
K$\alpha$ fluorescence photon at 1.74 keV. In this case, the residual energy
is $E_e = E - 1.74$\,keV. Otherwise, we have: $E_e = E$.

The newly formed charge cloud has an initial 1\,$\sigma$ radius of
\begin{equation}
r_i = 0.0044~ E_e^{1.75}\,\mu{\rm m}
\end{equation}
assuming that the spatial distribution of the charge follows a Gaussian
(Fitting et al. 1977). This charge cloud is collected in the buried channel in
the depletion region after spreading in the bulk of the detector.  The buried
channel is defined as the potential well where the photo-generated charge will
be collected and stored in a given CCD pixel between each readout. For the
CCD-22, the buried channel covers the entire pixel size. The charge cloud may
spread into more than one pixel, depending on its position with respect to the
electrode structure and its depth, which is a function of the value of the
linear absorption coefficients at the energy of the incoming photons ($E$) as
follows:
\begin{equation}
z = - \frac{1}{\mu_z(E)}~{\rm ln}(P+\epsilon)
\label{z}
\end{equation}
where $z$, $\mu_z(E)$ and $P$ are the depth of interaction of the incoming
photons, the linear absorption coefficient and a number uniformly and randomly
distributed between 0 and 1, respectively. A small offset $\epsilon\ll 1$ is
introduced to avoid any singularity when $P=0$.

The CCD pixel depth is divided into three regions:
\smallskip

- \underline{The depletion region} where the charges experience the full
effect of the electric field ($\mathbf{E}$). The radius of the charge cloud is
modified as it drifts to the buried channel. The contribution to the charge
cloud radius due to the spreading in the depletion region is given by:
\begin{equation}
r_d = \sqrt{\frac{2 k T~\epsilon_s}{e^2~N_a}~{\rm
ln}\left(\frac{z_d+\epsilon}{z_d-z+\epsilon}\right) }
\label{rd}
\end{equation}
where $z_d$ is the thickness of the depletion depth, $N_a$ -- the silicon
doping concentration ($N_a = 8\times 10^{12}$ cm$^{-3}$), $\epsilon_s$ -- the
Si permittivity ($\epsilon_s= 1.044 \times 10^{-12}$ F cm$^{-1}$), $k$ -- the
Boltzmann constant and $T$ -- the temperature. We use a temperature of
$-60^\circ$C in our code, which corresponds to the average CCD temperature
in-orbit (see Section 3.2.3). CCD temperature fluctuations by $\pm 10^\circ$C
do not change dramatically the $r_d$-values computed using Eq.~\ref{rd}
(i.e. by less than 2\%).  Equation.~\ref{rd} is undefined when
$z=z_d$. However, this singularity is not physical, so we introduced a small
offset $\epsilon = 0.002 \ll 1$ to avoid this.  All the charge reaching the
depletion region is collected in the buried channel.

\smallskip

- \underline{The field-free region} where the charge no longer feels the
effect of the electric field (i.e. a region where $\mathbf{E}\sim 0$). In this
region, the charge experiences diffusion, recombination and trapping in the
device lattice, so that the charge cloud can spread into several adjacent
pixels. The contribution to the charge cloud radius, due to spreading in the
field-free region, is given by (Janesick et al. 1985):
\begin{equation}
r_{\it ff} = \frac{z_{\it ff}}{2}~\sqrt{1-\left(\frac{z_2}{
    z_{\it ff}}\right)^2}
\label{rff}
\end{equation}
where $z_{\it ff}$ is the thickness of the field-free region and $z_2=z - z_d
    > 0$ is the depth of the interaction in the field-free region. Assuming
    that the recombination losses can be neglected, the charge reaching the
    depletion layer boundary is given by (Hopkinson 1983):
\begin{equation}
q_{\it ff} = Q_0 \frac{{\rm
cosh}\left(\frac{z_{\it ff}-(z-z_d)}{L_{\it ff}}\right)}{{\rm
cosh}\left(\frac{z_{\it ff}}{L_{\it ff}}\right)}
\label{qff}
\end{equation}
where $L_{\it ff}$ is the value of the diffusion coefficient in the field-free
region ($L_{\it ff}\sim 4000$\,$\mu$m). Equations.~\ref{rff} and~\ref{qff}
initially used to describe the diffusion in the field-free region have been
modified to follow the formalism described in Section 3 in Pavlov \& Nousek
(1999), because the assumption that the radial distribution of the charge is
Gaussian is no longer valid in this region (e.g. Pavlov \& Nousek 1999; see
also Section\,3.3.3).

\smallskip

- \underline{The substrate region} in which most of the charge produced
is lost by diffusion and recombination, because the electron diffusion
length is much shorter than in the two other regions. The contribution to the
charge cloud radius due to spreading in the substrate region is given by
(McCarthy et al. 1995)
\begin{equation}
r_s \sim \frac{L_s}{2.2}~\sqrt{1 - \left(\frac{z_s - z_3}{L_s}\right)^2}
\label{sub}
\end{equation}
where $z_s$ is the thickness of the substrate, $L_s$ -- the diffusion length in
the substrate which is much shorter than in the field-free region ($L_s =
10\,\mu$m for $N_a \sim 10^{18}$ cm$^{-3}$) and $z_3 = z - z_d - z_{\it ff}$ is
the depth of the interaction in the substrate.  The diffusion length is
replaced by $z_s$ when using Eq.~\ref{sub} to avoid having $(z_s-z_3) > L_s$
(Short et al. 2002).

The total 1\,$\sigma$ radius of the charge cloud reaching the buried channel
is then given by the quadratic summation of the relevant radii assuming that
the charge cloud profile due to radial diffusion is normally distributed:
\begin{equation}
r = \sqrt{r_i^2 + r_d^2 + r_{\it ff}^2 + r_s^2} \simeq \sqrt{r_d^2 + r_{\it ff}^2 + r_s^2}
\end{equation}

The contribution of $r_i$ can be neglected with respect to that from the
spreading radii $r_d$, $r_{\it ff}$ and $r_s$.

Figure.~\ref{fig_pixel} gives the thicknesses of these different layers as
used in the CCD response model. The thickness of the depletion depth is an
input parameter in our model. From quantum efficiency (QE) measurements
performed at the Leicester calibration facility, we demonstrated that a
27$\mu$m-thick depletion depth matches the on-ground calibration data using a
substrate voltage of $V_{\it ss}=0$\,V, as initially set in flight.

\subsubsection{Event recognition process }
\label{grade}

\begin{figure*}
\begin{center}
\includegraphics[width=14cm,height=14cm]{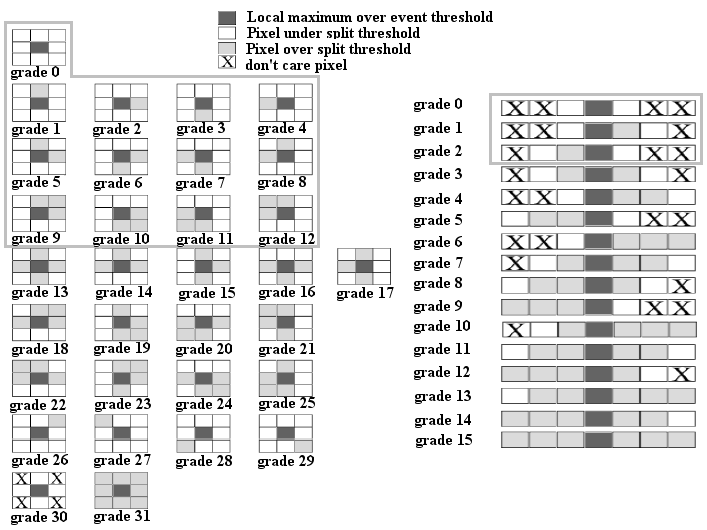}
\end{center}
\caption[]{List of the XRT event grades: (left) for PC mode, a $3\times 3$
  pixel matrix centred on the pixel containing the most charge is used to
  define the event grade, while for WT mode (right) a 7 pixel vector is
  used. The grades 0-12 in PC mode and the grades 0-2 in WT mode are
  considered as valid events due to X-rays as shown by the grey boxes in
  the Figure. }
\label{fig_grade}
\end{figure*}

The spreading of the charge cloud over several pixels implies that it is
essential to use an event grade recognition process to retrieve the right
energy of the incoming photons.  We use the same event grade recognition
process as implemented in the ground analysis software ({\scriptsize XRTDAS}
software package\,\footnote{{\scriptsize
http://swift.gsfc.nasa.gov/docs/swift/analysis/xrt$_{-}$swguide$_{-}$v1$_{-}$2.pdf}};
Capalbi et al. 2005).

An event is formed when a pixel has an analogue-to-digital converted (ADC)
charge greater than the event threshold (80\,DN\,\footnote{Digital
number}). This threshold is used to avoid pixels containing only noise from
swamping the telemetry.  In PC mode, the event data are telemetered as a
$3\times 3$ pixel matrix centred on the pixel with the most charge. Only the
surrounding pixels for which the charge is above the split event threshold
(40\,DN) are considered in the grade mapping, the other pixels being
discarded. The split event threshold was fixed at 40\,DN to minimise the
increase of the sub-threshold charge losses which result in a wing on the
low-energy side of the line profile, and hence to minimise the degradation of
the energy resolution (see Sections~\ref{phys} and \ref{shoulder}).

Since the WT mode only offers 1-D spatial imaging information (see Section~1),
it is impossible to use a $3\times 3$ pixel matrix in WT mode to classify the
X-rays events. Instead, a seven pixel vector centred on the pixel with
the most charge is used. In WT mode, the central and split event thresholds
are both set at 80 DN. Figure~\ref{fig_grade} shows the list of the 32 and 15
event grades which can be formed in PC and WT modes, respectively. The PC
grades are similar to those defined for the {\it XMM-Newton} MOS data (Burrows
et al. 2005).

To eliminate events due to charged particles, and to obtain good energy
resolution, we consider grades 0-12 in PC mode and grades 0-2 in WT mode
as valid X-ray events. From Fig.~\ref{fig_grade}, the WT grade 0 includes the
PC grades 0, 1 and 3 (i.e. the mono-pixel and vertical split events in PC
modes), while the WT grades 0-2 includes the good PC grades 0-12 as well as
possible higher PC grades 15, 17, 19, 21, 23, 25-29.

The pre-launch RMFs (v006) were released for three grade selections (0, 0-4 and
0-12) in PC mode and two grade selections (0 and 0-2) in WT mode. This was to
offer the user the choice of higher spectral resolution at the cost of lower
effective area. The decision was made after launch not to upgrade the PC
grade 0-4 RMFs after the release of the v007 response files, since the PC
grade 0-12 RMF offers a higher quantum efficiency at high energy and its
calibration was sufficiently good.

\subsubsection{X-ray spectrum from monochromatic radiation}
\label{phys}

The X-ray event energy spectrum resulting from monochromatic radiation with an
energy $E$ significantly differs from a simple Gaussian. It consists of
multiple components: a Gaussian photo-peak with a shoulder on the low energy
side, a shelf extending to low energies, and, at the very lowest energies, the
high-energy side of a noise peak above the on-board central event threshold
(see Section~\ref{grade}). For photon energies above the Si K-shell edge
(1.839 keV in the silicon bulk) two additional features appear: an escape peak
of energy $E-E_{\it Si}$, and a Si K$\alpha$ fluorescence peak at $E_{\it Si}
= 1.74$ keV.

The DN value of the minimum energy threshold ($thres$) in the spectra depends
on the mode, the grade selection and the settings for the event threshold and
the split event threshold in the ground processing code. The default threshold
configuration for PC mode is 80\,DN and 40\,DN, respectively, and for WT mode
both are set to 80\,DN. Therefore, in PC mode, the default minimum energy
threshold for grade 0 events is 80\,DN while it is set at ($80+40$)\,DN for
double split events, ($80+(2\times 40$))\,DN for triple split events and
($80+(3\times 40$))\,DN for quadruple split events. In WT mode, the minimum
energy threshold is set at 80\,DN and $80+80$\,DN for grade 0 and grades 1 \&
2 events, respectively.  The $thres$-values once expressed in units of eV
[\emph{i.e.} $thres \times C_0\times G$ where $C_0$ and $G~(=10)$ are the
multiplicative DN to PHA (pulse height amplitude) gain factor and the global
PHA to PI (pulse invariant) gain factor, respectively - see Eq.~\ref{PI}]
slightly increased over time, since the gain coefficient $C_0$ has increased
by about 3\% from launch to June 2007 (see Fig.~\ref{fig_CTI2} and
Section~\ref{CTI}).

\begin{figure}[h]
\begin{center}
\includegraphics[width=6.5cm,angle=-90]{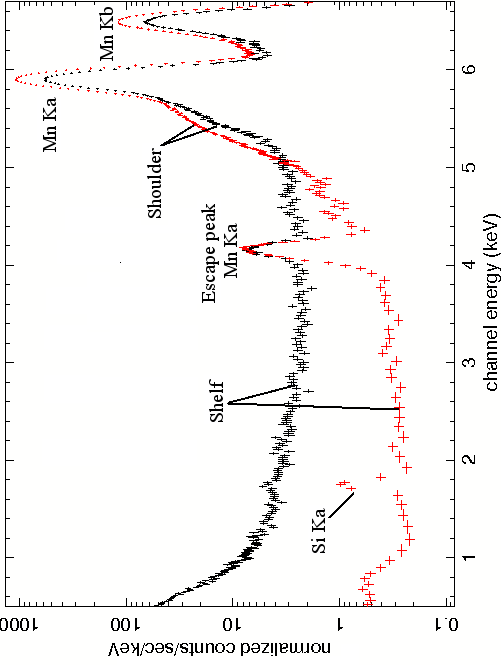}
\end{center}
\caption[]{Spectra from the Mn K$\alpha$ and K$\beta$ lines (5.9 and 6.5 keV)
  from an on-board $^{55}$Fe radioactive source in WT mode (black crosses -
  grades 0-2) and PC mode (red crosses - grades 0-12).}
\label{fig_spec_WT}
\end{figure}

At low energy, a large fraction of the photons interact at the interface
between the SiO$_2$ layer placed on top of the silicon bulk and the active
silicon bulk itself. Short et al. (2002) measured the lost fraction of the
total charge formed at that interface.  This energy loss results in a
low-energy wing in the line profile as well as a shelf extending to low
energy. The effects of the surface charge losses start to become very
significant below 0.5 keV, since the photo-peak disappears almost completely
and the resulting line profile is shifted to lower energies (see Short et
al. 2002).  The authors proposed that the surface loss effect may be due to
the charge-state of the surface oxide so that the oxide layer is charged
enough in the etched regions (open electrode areas shown in white in the
bottom left panel in Fig.~\ref{fig_pixel}) to cause a local turn-over of the
potential (see Fig. 7 in Short et al. 2002). In this case, the charges formed
near the oxide layer will move to the surface rather than to the buried
channel. The surface charge losses are a function of the depth of interaction
and the energy.  To model this double dependency, the XRT energy range is
divided into 12 energy bands ($\Delta E_{i=1,12}$). In each energy band, the
surface charge losses were initially defined by a set of linear functions
$f(\Delta E_{i=1,12},z)$ depending on the depth of interaction $z$ inside the
silicon bulk, the coefficients of the function $f(\Delta E_{i=1,12},z)$ being
empirically derived from spectroscopic measurements performed at different
energies at the Leicester calibration facility using radioactive elements.

At higher energies, when the photons interact more deeply in the CCD (above 2
keV), other processes start to become more important in the production of the
shoulder and the shelf: (i) sub-threshold losses (see Section~\ref{shoulder});
(ii) diffusion, recombination and trapping in the bulk of the detector; (iii)
inhomogeneity of the electric field in the depletion depth; these act in
addition to the surface losses.  The exact shape of the shoulder and the shelf
depend on the readout mode (see Fig.~\ref{fig_spec_WT} and the top panel in
Fig.~\ref{fig_low-E_ground}).

%################################
\subsection{Performance of the pre-launch RMFs}

The XRT CCD was calibrated using the EPIC calibration facility at the
University of Leicester, using sixteen discrete energies covering the 0.3-10
keV energy range. These measurements were used to interpolate and generate the
redistribution component of the XRT RMFs. The RMFs were initially calibrated
for a value of the substrate voltage of $V_{\it ss}=0$ V, which was the value
used from launch to 2007 August 30 (see Section~\ref{Vss}).

\begin{figure}[h]
\begin{center}
\begin{tabular}{c}
\includegraphics[width=7cm]{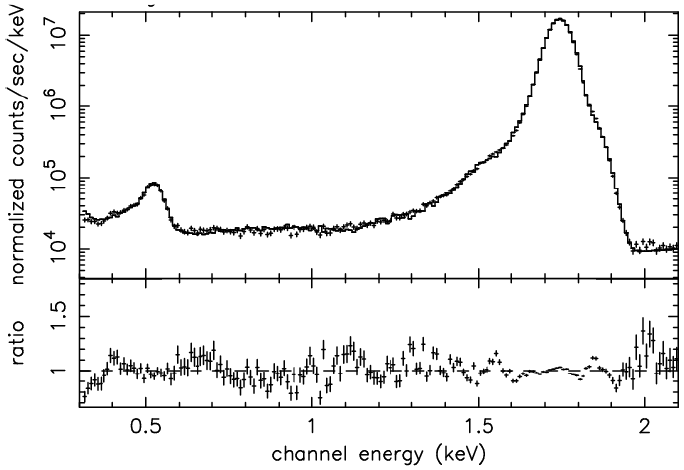}\\
\includegraphics[width=5.cm,angle=-90]{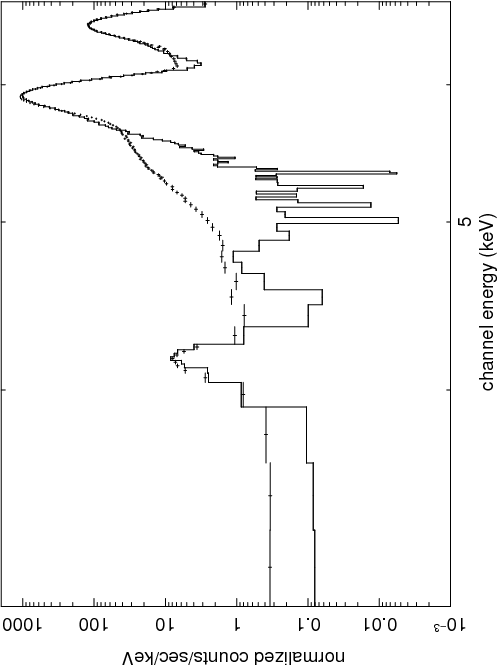}\\
\end{tabular}
\end{center}
\caption[]{(Top panel) PC grade 0 spectrum measured at the Leicester
  calibration facility from the Si K$\alpha$ and Si K$\beta$ lines (1.74 and
  1.84 keV) with background O K$\alpha$ (0.525 keV) and bremsstrahlung
  continuum, together with pre-launch RMF v006. (Bottom panel) PC grade 0-12
  spectrum measured at the Leicester calibration facility Mn K$\alpha$ and
  K$\beta$ lines (5.89 and 6.5 keV) from an on-board $^{55}$Fe source together
  with the pre-launch PC RMF v006. The low-energy shoulder and shelf are not
  fully reproduced by the model.}
\label{fig_low-E_ground}
\end{figure}

The pre-launch RMFs were computed to model the spectral response in the
central region of the CCD, i.e. a window of $200\times 200$ pixels (a field of
view of about $7.9\times 7.9$ arc-minutes$^2$). This area also corresponds to
the area on the CCD where most of the GRB X-ray afterglows are located after
the spacecraft slews.

Although the pre-launch RMFs showed good agreement between the calibration
data and the model in many circumstances, as displayed in the top panel of
Fig.~\ref{fig_low-E_ground}, there were still some modelling issues. The
low-energy shoulder of the high-energy ($E > 2-3$ keV) line profiles and
the shelf needed improvement (see the bottom panel in
Fig.~\ref{fig_low-E_ground}).

%################################
\section{In-flight calibration}
\label{cal}

\subsection{Calibration overview}

\begin{figure}[h]
\begin{center}
\includegraphics[width=8.cm]{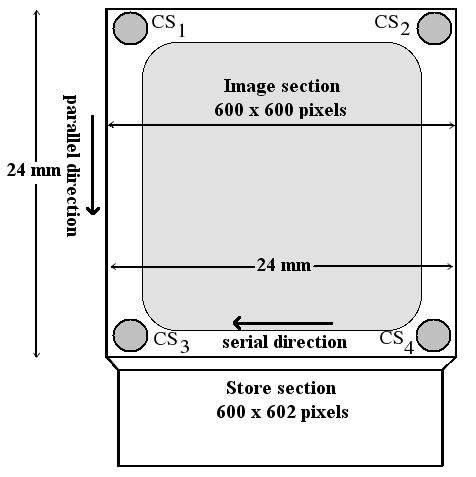}
\caption{CCD-22 simplified schematic diagram. The four $^{55}$Fe corner
sources permanently illuminating areas of the CCD outside the field of view
are shown as grey circles.}
\label{fig_CS}
\end{center}
\end{figure}

Regular in-flight spectroscopic calibration observations of a set of
well-known celestial objects are performed (with a frequency of about 6
months) in order to monitor changes in the response (see Table~\ref{tab1}).
The fraction of time spent on calibration since the launch is $\sim 7\%$ of
the total in-orbit time. In addition, we make use of four $^{55}$Fe
calibration sources which are located in each corner of the CCD and which
permanently illuminate a small fraction of the CCD area outside the XRT field
of view as shown in Fig.~\ref{fig_CS}. Before the focal plane camera assembly
(FPCA) door was opened, we also made use of an $^{55}$Fe calibration source
located on the inside of the door, which illuminated the entire CCD imaging
area.

\begin{table*}
\caption{Summary of the in-flight calibration targets used since the launch on
  2004 November 20.  }
\label{tab1}
\begin{center}     
\begin{tabular}{|l|l|l|l|l|}
\hline
Object$^\ddagger$  & Type & Mode & Purpose &  Exposure (ks) \\
\hline
RXJ 1856.5-3754$^{x,c,s}$ & Neutron star & PC/WT & Low energy response & 81/70 (31/40)$^\dagger$ \\
\hline
PKS 0745-19 & Cluster of & PC & Effective area & 63 (20)\\
            & galaxies  &    &                &    \\
\hline
2E 0102-723$^{x,c,s}$ & Line-rich SNR & PC & Gain, energy resolution and    & 119 (28)\\
            &    & WT & shoulder    & 76 (26) \\
\hline
Cas A$^{x,c,s}$  & Line-rich SNR & PC & Energy scale offset, gain, shoulder, &
235  (67)\\
            &    & WT & CTI and energy resolution  &  62 (39)  \\
\hline
3C 273$^{*/x,r,c,s}$      & Quasar & WT & Effective area & 18  \\
\hline
PSR 0540-69$^x$  & Pulsar & PC/WT & Effective area    & 56/26 (25 in PC)\\
\hline
PKS 2155-304$^*/x$ & Blazar & PC/WT & Effective area  & 14/13 (5/9)\\
           
\hline
NGC 7172$^{*/x}$    & Seyfert 2 & PC & Redistribution & 15  \\
\hline
G21.5$^x$        & Featureless SNR    & PC & High-energy shelf  & 43 (78 in PC
\& 77 in WT) \\
\hline
Mkn 421$^*$      & Blazar & WT & Effective area  & 33 (10) \\
\hline
Crab$^{x,c,s,r,i}$      & Pulsar/SNR & WT & Effective area  & 46 (5) \\
\hline
\end{tabular}
    \begin{list}{}{}
      \item $^\ddagger$ The letter indices $x,\,c,\,s,\,r,\,i$ correspond to
  sources observed by other X-/Gamma-ray instruments for calibration purposes:
  {\it XMM-Newton}; {\it Chandra}; {\it Suzaku}; {\it RXTE}; {\it INTEGRAL},
  respectively.

\item $^\dagger$ The numbers in parentheses correspond to the exposure
  time collected from September 2007 to the end of December 2007 and dedicated
  to the re-calibration of the XRT after the substrate voltage change (see
  Section~\ref{Vss}), while the other numbers correspond to the calibration
  data collected prior to the substrate voltage change on 2007 August 30.

      \item $^*$ This symbol corresponds to the sources for which simultaneous
  calibration observations were performed with the {\it XMM-Newton} EPIC
  cameras.

    \end{list}
\end{center}
\end{table*}

Many of our calibration targets are also observed by other X-ray observatories
such as {\it XMM-Newton}, {\it Chandra}, {\it Suzaku} and {\it RXTE}, enabling
us to perform cross-calibration campaigns. So far, five cross-calibration
campaigns with the {\it XMM-Newton} EPIC cameras on variable sources have been
made (see Table~\ref{tab1} and Section~\ref{perfo}). We also use the set of
stable calibration sources observed by other X-ray instruments to compare and
improve the performance of our spectral response (see Table~\ref{tab1} and
Section 4.1).

\subsection{Energy scale offsets and origins}
\label{offset}

Before describing the post-launch improvements made in our CCD response model,
we address an important issue related to energy scale offsets, since they can
lead to misinterpretation of the data and hence strongly affect the modelling
of the response, especially around the instrumental edges.  Below, we discuss
four different causes resulting in energy scale offsets and describe the
solutions which were found to correct the XRT energy scale.

\begin{table*}
\caption{Summary of the releases of the XRT post-launch gain files. }
\label{tab_gain}
\begin{center}
\begin{tabular}{|c|c|l|c|}
\hline
Release  & Release  & Main improvements in the gain & Text section \\
number$^\dagger$  & date  &  & number\\
\hline
\hline

05  & 2005 April 5  &  Introduction of a temperature-dependent gain file & 3.2.3  \\
\hline

06  &  2005 October 28  &  Update of the gain coefficients over time & - \\
\hline
07$^*$  & 2007 July 9  &  Update of the CTI values over time in WT \& PC
mode & 3.2.2\\
    &    &  &\\
    &    &  Update of the gain coefficients in WT \& PC
    modes after the substrate & 4.3\\
    &    & voltage ($V_{ss}$) change from 0\,V to 6\,V on 2007 August 30 &\\
\hline

08 & 2008 June 25 & Introduction of an offset of $C_3=17.6$ eV to restore the
WT energy scale & 3.2.4\\
\hline
\end{tabular}
    \begin{list}{}{}
      \item $^\dagger$ The
  v003-v004 gain files correspond to the pre-launch gain files.
      \item $^*$ The gain file has now added a suffix `s0' or `s6' to indicate
  the appropriate substrate voltage ($V_{ss}$).  The `s0' files are valid from
  the start of the mission until when the substrate voltage was raised to
  $V_{ss}=6$\,V on 2007 August 30.
    \end{list}
\end{center}
\end{table*}

\subsubsection{Evaluation of the bias level}
\label{bias}

\begin{figure}[h]
\begin{center}
\begin{tabular}{c}
\includegraphics[width=6cm]{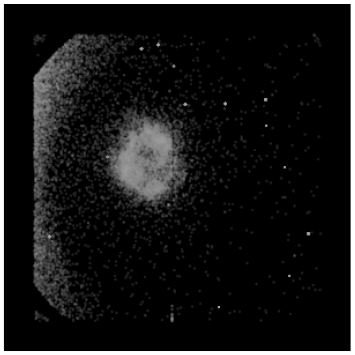} \\
\includegraphics[width=8cm]{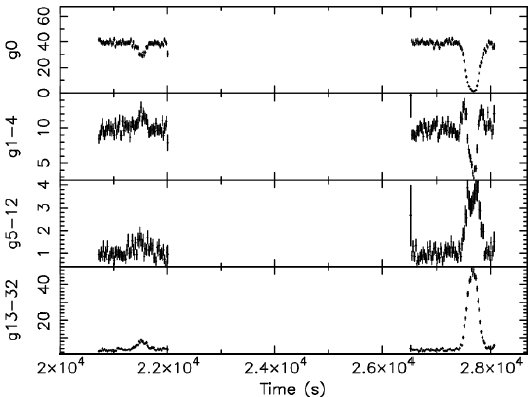} \\
\end{tabular}
\caption{(Top panel) Detector coordinate image showing the SNR Cas A when
  scattered optical light from the sunlit Earth is present on the detector
  (visible on the left side of the detector).  (Bottom panel) PC mode
  light-curve of the SNR Cas A in the 0.3-10 keV energy range for different
  snapshots and grade selections: 0, 1-4, 5-12 and 13-31. The light-curve was
  obtained before bias correction. The drop of the count rate in the PC grade
  0 light-curve is due to the presence of scattered optical light from the
  sunlit Earth, which induces a grade migration from good grades (0-12) to
  higher grades.}
\label{fig_BE}
\end{center}
\end{figure}

In addition to thermal noise, each pixel charge will carry with it a fixed
zero-point offset DN value, a random readout noise from the amplifier and the
noise from camera electronics. The zero-point offset defines the bias level
measured on the CCD. The amplifier readout noise and the electronic noise,
however, affect the determination of the bias level, but we minimize those
contributions by averaging multiple bias measurements. In any case, the noise
contributions should be much less than the measured offsets.  In orbit, bias
frames in PC mode and bias rows in WT mode are taken during each slew of the
spacecraft to a new target before the beginning of an observation.  For PC
mode, the bias is computed on a pixel-by-pixel basis. Five $600\times 600$
pixel bias frames are taken, averaging each into the existing bias map using a
``running-mean'' algorithm that sets each pixel $(x,y)$ to a new mean value
computed as follows:
$${\rm MeanPix}_{i+1}(x,y)=\frac{(N-1)}{N}{\rm MeanPix}_{i}(x,y) + \frac{{\rm
Pix}_i(x,y)}{N}$$ where $N=3$ is the running mean length and $i=1,...,~5$ is
the frame count.  For WT mode, the bias row is a vector of bias values, one
for each column in the WT mode window (i.e. 200 columns).  The bias row vector
is computed over one image of 600 rows.  A ``running-mean'' approximation is
computed for each pixel $x$ in each row according to the following equation:
$${\rm MeanPix}_{i+1}(x)=\frac{(N-1)}{N}{\rm MeanPix}_{i}(x) + \frac{{\rm
Pix}_i(x)}{N}$$ with $N=3$ and $i$ corresponding to the row. The bias
information is then used to correct on-board the subsequent frames of data
during the XRT observations.

\begin{figure}[h]
\begin{center}
\begin{tabular}{c}
\includegraphics[width=7.cm,height=6cm]{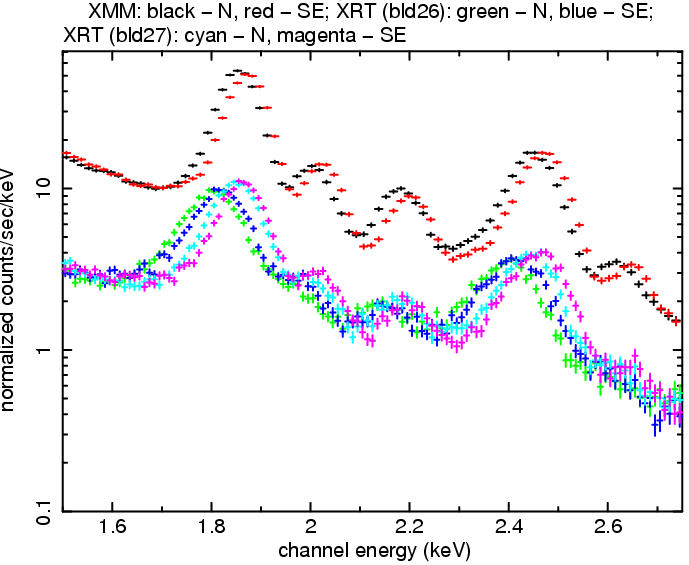} \\ 
\includegraphics[width=7.cm,height=6cm]{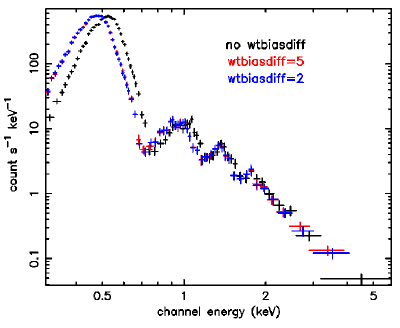}\\
\end{tabular}
\caption{(Top panel) Comparison of the event energy spectra of the Si and S
  lines in the North (N) and South East (SE) knots of the SNR Cas A as
  observed by the {\it XMM-Newton} MOS cameras (N: black; SE: red) and the
  {\it Swift}-XRT (N: green, cyan; SE: blue, magenta). The green and blue
  crosses correspond to XRT/PC grade 0 data for which the bias was
  contaminated by optical light from the sunlit Earth. The data were processed
  with the v2.6 XRT software which does not correct for bias contamination. In
  this case, an energy scale offset is observed when compared with the {\it
  XMM-Newton} MOS spectra. The cyan and magenta crosses correspond to the same
  data processed with the XRT software including the task {\scriptsize
  XRTPCBIAS}, which corrects the data.  (Bottom panel) {\it Swift}-XRT WT
  grade 0-2 spectrum of RS Ophiuchi: ({\it black}) the data not corrected for
  the bias contamination and ({\it red and blue}) the data corrected using the
  modified task {\scriptsize XRTWTCORR}.}
\label{fig_scale}
\end{center}
\end{figure}

The bias level is mode dependent, and we have also seen observational evidence
that the bias level can significantly vary even during a single snapshot on a
celestial target (i.e. over $\sim 20$ minutes).  Variations of the measured
bias level can be due to changes in the CCD temperature following some sky
pointing directions. Shifts in the CCD bias level are also caused by roughly
sinusoidal temperature variations of $\pm3^\circ$C during each 95 minute
orbit. The measured bias level can also be contaminated by scattered optical
light from the sunlit Earth as shown in the top panel in Fig.~\ref{fig_BE}
(see Beardmore et al. 2007 for more details). Scattered optical light from the
sunlit Earth mainly affects PC mode because of its much longer frame
accumulation time than WT mode, and seems to occur when the instrument is
pointed near the sunlit Earth horizon.  Changes in the bias level in both WT
and PC modes result in energy scale offsets (see Fig.~\ref{fig_scale}).

In order to correct the bias level on the ground, and hence restore the energy
scale (see Fig.~\ref{fig_scale}), the task {\scriptsize XRTWTCORR}, for WT
mode, was enabled in the {\scriptsize XRTDAS} software package (Swift software
v.2.6). For PC mode, a new task {\scriptsize XRTPCBIAS} was developed and
released with Swift software v.2.7.

In addition to producing energy scale offsets, in some extreme cases scattered
optical light in PC mode can induce a significant grade migration from good
grades (grades 0-12) to higher rejected grades, resulting in an apparent loss
of counts in the light-curve (see the bottom panel in Fig.~\ref{fig_BE}). In
these extreme cases, the {\scriptsize XRTPCBIAS} task will not be able
completely to correct the bias level and a residual energy scale offset will
be still present in the spectra.

\subsubsection{Increase of charge transfer inefficiency}
\label{CTI}

\begin{figure}
\begin{center}
\begin{tabular}{c}
\includegraphics[width=5.2cm,angle=-90]{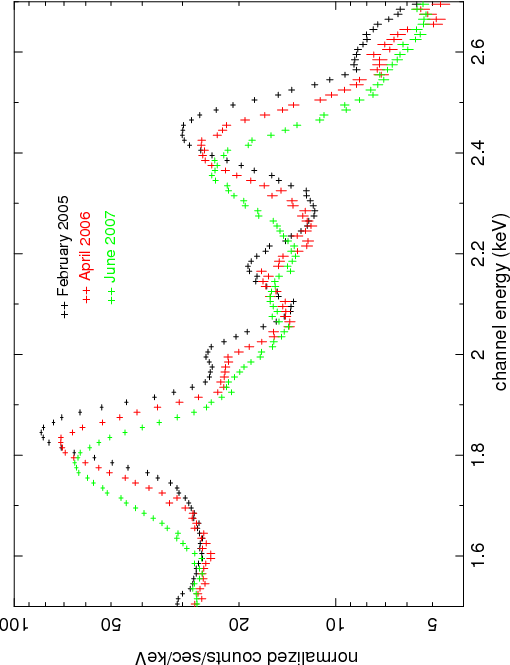}\\
\includegraphics[width=5.2cm,angle=-90]{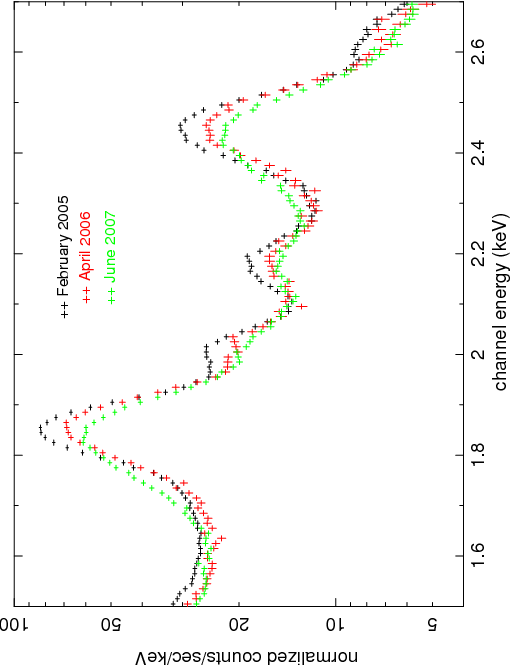}\\
\end{tabular}
\caption{(Top panel) WT grade 0-2 spectrum of the SNR Cas A in the energy band
  of the Si K$\alpha$ and S K$\alpha$ lines at different epochs since launch
  showing the degradation of the gain due to the CTI increase. The WT data
  were processed using the option {\scriptsize BIASDIFF} in the task
  {\scriptsize XRTWTCORR} correcting the bias level if necessary (see
  Section~\ref{bias}). The energy centroids of the Si K$\alpha$ and S
  K$\alpha$ lines show an energy shift of $-$50\,eV from February 2005 to June
  2007 (see Section~\ref{trap} for details about the line broadening). (Bottom
  panel) As above, but with the gain corrected for the effects of CTI.}
\label{fig_CTI}
\end{center}
\end{figure}

\begin{figure}
\begin{center}
\includegraphics[width=9cm]{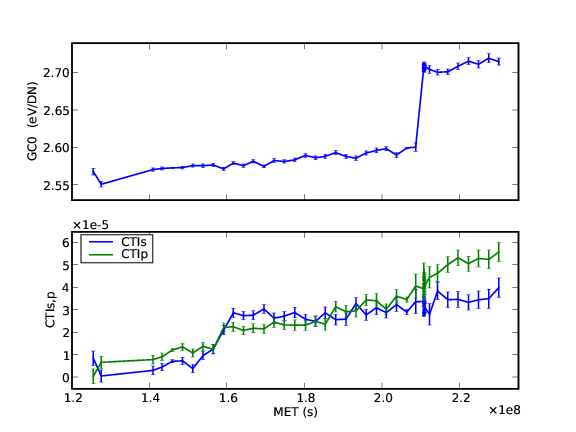}
\caption{(Top panel) Evolution of the gain $C_0$ at a CCD temperature of
  $-$65$^\circ$C. Note the jump in gain around MET (Mission Elapsed Time)
  $2.1\times 10^8$\,s due to the substrate voltage change from $V_{\it ss}
  =0$\,V to 6\,V (see Section~\ref{Vss}). (Bottom panel) Evolution of the
  serial (the blue line) and parallel (the green line) CTI values over
  time. Note the increase in CTI around MET $1.6\times 10^8$\,s (\emph{i.e.}
  January 2006). }
\label{fig_CTI2}
\end{center}
\end{figure}

CCD detectors provide good X-ray imaging and spectroscopic
performance. However, the increase of charge transfer inefficiency (CTI) over
time due to radiation damage is a fundamental limitation of CCD technology.
Due to the effects of CTI, charge is lost during the readout process so that
the remaining charge $Q$ reaching the output amplifier after $N_s$ transfers
in the serial direction and $N_p$ transfers in the parallel direction is
\begin{equation}
Q = Q_0 (1-{\rm CTI}_s)^{N_s} (1-{\rm CTI}_p)^{N_p}
\label{Q}
\end{equation}
where $Q_0$ is the initial charge and CTI$_{\it s,p}$ are the serial
and parallel CTI values, respectively.  The CTI increase in the imaging area,
the store-frame area and the serial register can then result in an energy
scale offset if the data are not corrected. As an example, Fig.~\ref{fig_CTI}
shows the WT grade 0-2 Cas A spectrum in the Si K$\alpha$ and S K$\alpha$ line
region at three different epochs when the data are not corrected for the
effects of CTI. An offset of $-$50\,eV is clearly visible between February 2005
and June 2007.

The XRT gain file takes into account the effect of CTI as follows:
\begin{equation}
{\rm PI} =\,\frac{{\rm PHA}\times (C_0 + C_1 \, x + C_2 \, y) + C_3
}{G}
\label{PI}
\end{equation}
where $x$ and $y$ correspond to the position of the event in the CCD imaging
area. $G=10$ is the global PHA to PI gain factor. $C_{0}$ is the
multiplicative DN to PHA gain factor, while $C_{1,2} = C_0\times
\mathrm{CTI}_{\it s,p}$ are the multiplicative serial and parallel CTI
correction factors, respectively (see Pagani et al. 2008 for more details).
$C_3$ is an additive offset correction factor which was set to 0 in both PC
and WT modes before launch.  $C_{0,1,2}$ are functions of the CCD temperature
(see Section~\ref{TEC}) and time. The CTI$_{\it s,p}$ values used to compute
the $C_{1,2}$ coefficients in the gain file are the same for all CCD
temperatures (see Section~\ref{TEC}).

Since launch, the evolution of serial and parallel CTI and the gain over time
have been monitored using the four corner calibration sources (see
Fig.~\ref{fig_CS}).  The gain $C_0$ is given by the gain of the bottom left
corner source CS$_3$ closest to the output amplifier, as this does not suffer
from CTI loss in the imaging area.  $C_0$ thus includes both the output
amplifier gain and the degradation in the gain caused by charge loss during
the frame-store transfer. CTI$_s$ is measured using corner source pair CS$_3$
and CS$_4$, while CTI$_p$ measured using corner source pair CS$_1$ and CS$_3$.
Figure~\ref{fig_CTI2} shows the evolution of the gain $C_0$ (from 2.529 just
after launch to 2.625 in June 2007) and the CTI$_{\it s,p}$ values over time.
Figure~\ref{fig_CTI2} shows a jump in gain $C_0$ around MET (Mission Elapsed
Time) $2.1\times 10^8$\,s due to the substrate voltage change from $V_{\it ss}
=0$\,V to 6\,V (see Section~\ref{Vss}). There also is a jump in CTI around MET
$1.6\times 10^8$\,s; the origin of this jump is however unknown. The
correction of the data from the effects of CTI is illustrated in the bottom
panel of Fig.~\ref{fig_CTI}. These CTI values were implemented in the PC and
WT gain files from version 007 and onwards (see Table~\ref{tab_gain}).

\subsubsection{The effect of temperature}
\label{TEC}

The XRT was designed to cool the CCD to a nominal operating temperature of
-100$^{\circ}$C using a thermo-electronic cooler (TEC). However, the TEC power
supply system apparently failed shortly after launch, and the XRT has to rely
on passive cooling via a heat pipe and radiator, in combination with enhanced
management of the spacecraft orientation to reduce the radiator view of the
sunlit Earth.  The XRT is now operated with CCD temperatures varying between
-75 to $-52^{\circ}$C (see Kennea et al. 2007 for more details), though new
GRBs occasionally cause it to exceed $-52^\circ$C.

\begin{figure}[h]
\begin{center}
\begin{tabular}{c}
\includegraphics[width=6.5cm,angle=-90]{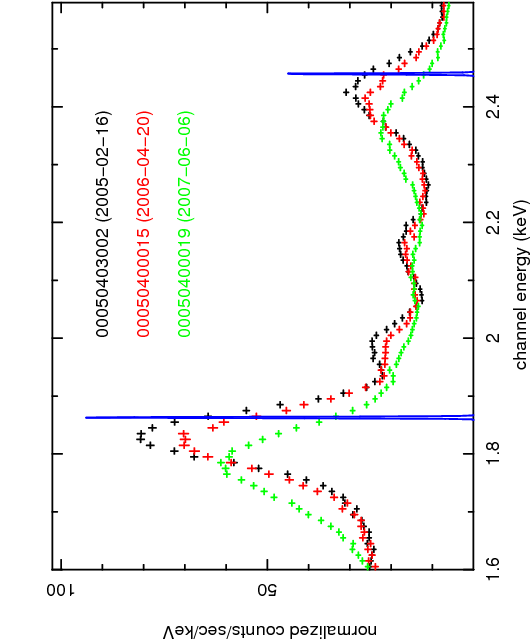} \\
\vspace{-0.3cm}
\includegraphics[width=6.5cm,angle=-90]{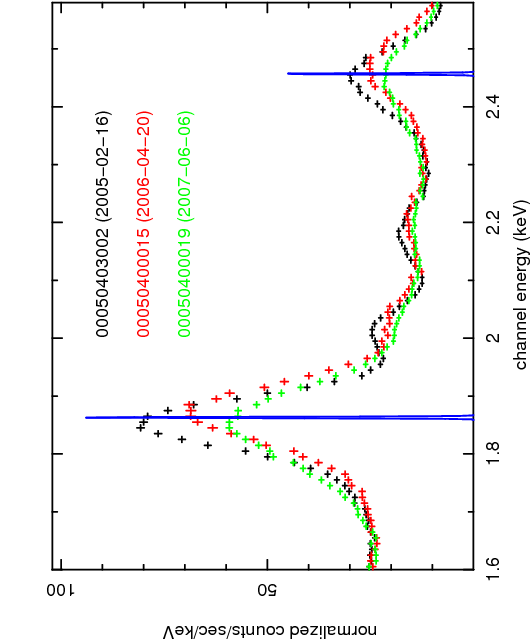} \\
\end{tabular}
\caption{ Spectra of the entire SNR Cas A in WT mode at different epochs:
  (black) 2005-02-16; (red) 2006-04-20; (green) 2007-06-06. The blue line
  indicates the expected energy centroid of the Si and S K$\alpha$ lines. (Top
  panel) Data processed using the v007 WT gain file. (Bottom panel) Data
  processed using the v008 WT gain file including an offset of 17.6 eV.}
\label{fig_WT-PC}
\end{center}
\end{figure}

Since the gain of the CCD output FET (Field Effect Transistor) is sensitive to
the temperature, the reduced temperature stability required the introduction
of a temperature dependency in the gain expression in order to restore the XRT
energy scale. In addition to the K$\alpha$ and K$\beta$ fluorescent lines of
the corner source data, we used the supernova remnant Cas A, because its
spectrum shows intense silicon (1.86 keV) and sulphur (2.45 keV) lines as well
as an iron (6.6 keV) line. The gain coefficient $C_{0}$ in Eq.~\ref{PI} was
computed at two CCD temperatures other than the $-100^\circ$C used during the
ground calibration: $-65^\circ$C and $-48^\circ$C using in orbit Cas A data.
The ground data processing software linearly interpolates between tables of
gain coefficients for these three different temperatures to determine the
correct value to use for a given
observation\,\footnote{http://swift.gsfc.nasa.gov/docs/heasarc/caldb/swift/docs/xrt/\\SWIFT-XRT-CALDB-04$_{-}$v2.pdf}. These
temperature-dependent values of $C_0$ were used to compute the coefficients
$C_{1,2}$ as well (see Section~\ref{CTI}). The linear slope characterizing the
gain change as a function of temperature from $-65^\circ$C to $-48^\circ$C is
0.00117 eV DN$^{-1}$ $^\circ$C$^{-1}$. This corresponds to a $\sim 7$ eV
variation in the energy scale at 1.5 keV for a 10$^\circ$C temperature
variation.  This temperature dependency on the gain was included in the v005
release of the PC and WT gain files (see Table~\ref{tab_gain}).

\subsubsection{Comparison of the WT and PC energy scale}
\label{WTscale}

We found that even after correcting the data for the various sources of energy
scale offsets discussed in previous sections, there was still a slight
systematic mismatch between the WT and PC energy scales (see the top panel in
Fig.~\ref{fig_WT-PC}). We estimated the systematic offset to be $17.6$\,eV,
based on the comparison of the line centroids between WT and PC Cas A spectra,
as well as the use of the XMM-RGS model to fit the XRT/WT spectra of the SNR
E0102-723 using the gain command within Xspec (Arnaud 1996). This energy
shift appears to be independent of time, temperature and energy. The setting
of the offset term $C_3$ in Eq.~\ref{PI} at 17.6 eV in the WT gain file was
shown significantly to improve the WT energy scale (see the bottom panel in
Fig.~\ref{fig_WT-PC}). This new v008 WT gain file was released on 2008 June 25
(see Table~\ref{tab_gain}). The origin of the energy offset in WT mode is
unclear and still under investigation.

\subsection{Improvements to the response model}
\label{impro}

We discuss in this section the improvements made to the CCD response model
based on experiences from in-flight calibration. Table~\ref{tab_release}
summarises the improvements made to the response files and the corresponding
release number.

\begin{table*}
\caption{Summary of the releases of the XRT post-launch RMFs.   }
\label{tab_release}
\begin{center}
\begin{tabular}{|c|c|l|c|}
\hline
Release number$^\dagger$  & Release date  & Main improvements in the RMFs & Text section
number\\
\hline

07  & 2005 April 5  &  {\bf Line shoulder:} ad-hoc increase of threshold to favour & 3.3.3  \\
    &    &  sub-threshold losses in PC \& WT mode & \\
\hline

08  &  2006 April 24  &  {\bf Low-energy response:} change in the description
    the  & 3.3.1 \\
    &    &  surface loss function in PC \& WT mode  &    \\
\hline
09$^\ddagger$  & 2007 May 31  &  {\bf Line shoulder:} New description of the charge
    cloud in & 3.3.3\\
    &    &  the field-free region for high-energy photons ($E > 2$ keV) &\\
    &    &  in PC mode &\\
    &    &  &\\
    &    &  {\bf Shelf:} Rescaling of the shelf in both PC \& WT mode & 3.3.2\\
\hline

11$^*$  & 2008 June 25 & {\bf Loss function:} New description of the surface loss function & 3.3.4\\
      &            & in the 1-2 keV energy range in WT mode & \\
\hline
\end{tabular}
    \begin{list}{}{}
      \item $^\dagger$ The v006 response files correspond to the pre-launch
  response files.
      \item $^\ddagger$ The v009 response files were renamed as v010 response
  files when the substrate voltage was raised from 0 to 6\,V on 2007 August 30
  (see Section 4.3) following an update of the ground software.
      \item $^*$
      http://swift.gsfc.nasa.gov/docs/heasarc/caldb/swift/docs/xrt/SWIFT-XRT-CALDB-09$_{-}$v11.pdf
      (Godet et al. 2008)
    \end{list}
\end{center}
\end{table*}

\subsubsection{The low-energy response}
\label{low_E}

\begin{figure}[h]
\begin{center}
\hspace{0.3cm}\includegraphics[width=6.cm,angle=-90]{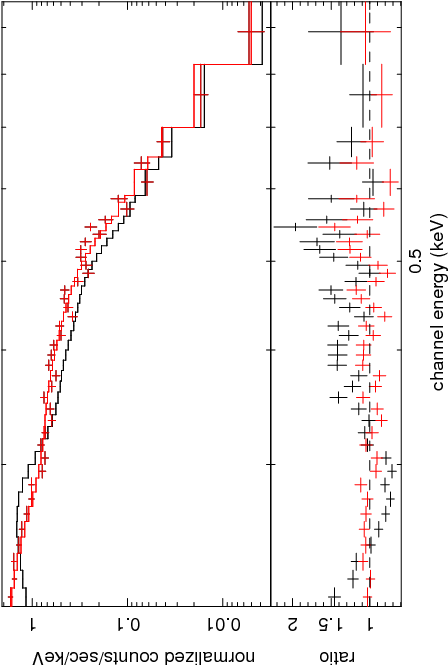}
\end{center}
\caption[]{Comparison of the low-energy response between v007 (black) and v008
  (red) RMFs using the spectrum of the soft neutron star RX J1856.4-3754 in PC
  mode for grade 0 events. We used a {\scriptsize
  CONST*WABS*(BBODYRAD+BBODYRAD)} model with the spectral parameters fixed to
  the values given in Beuermann et al. (2006).  }
\label{fig_low}
\end{figure}

Shortly after launch, it became apparent that the low energy ($E < 0.5$ keV)
response model could be improved.  Figure~\ref{fig_low} shows that the PC grade
0 spectrum of the soft neutron star RX\,J1856.4-3754 (the black curve) is
poorly fitted when the v007 RMF is used. A {\scriptsize
CONST*WABS*(BBODYRAD+BBODYRAD)} model with the spectral parameters fixed to
the values given in Beuermann et al. (2006) (\emph{i.e.}  $N_H=1.1\times
10^{20}\,\rm cm^{-2}$, $kT_1 = 62.8$ eV and $kT_2 = 32.3$ eV) was used for the
fit. The lowest temperature black-body component has a minor impact in the XRT
energy range. It was introduced by Beuermann et al. (2006) to fit the EUVE
data as well as the Chandra data.  Below 0.3 keV, the modelled line profile
and its energy centroid are strongly dependent on the estimate of the charge
losses at the CCD surface. As explained in Section 2.1.3, these charge losses
are a function of energy and location of interaction of incoming photons. To
model these charge losses better, we used the semi-empirical formalism
described in Popp et al. (2000) to describe the loss function $f(\Delta
E_{i=1,12},z)$ in our CCD response model as follows:

\[f(\Delta E_{i=1,12},z) = \left\{
\begin{array}{ll}
0, &  z \le 0~{\rm or}~ z \ge \Delta z_{\it CCD} \\

f_0 + A~\left(\frac{z}{l}\right)^c, &  0 \le z < l  \\

1 - B~{\rm e}^{-\frac{(z-l)}{\tau}}, &  l \le z < \Delta z_{\it CCD}  \\

\end{array}
\right. \] where $f_0$, $c$, $l$ and $\tau$ are free parameters which were
estimated initially by fitting the previous loss functions in each energy
band $\Delta E_{i=1,12}$ as defined in Section 2.1.3, and then by using
an iterative process of RMF generation and fits of spectra of soft neutron
stars and ground calibration data. $\Delta z_{\it CCD}$ is the CCD thickness
($\Delta z_{\it CCD} = 280\,\mu$m).  The parameters $A$ and $B$ were derived
using $f(\Delta E_{i=1,12},z=l)$ and $f'(\Delta E_{i=1,12},z=l)\equiv
\frac{df}{dz}|_{z=l}$.  This new description gives better results as shown in
Fig.~\ref{fig_low} (the red curve). This description was included in the v008
release of the PC and WT RMFs.

\begin{figure*}
\begin{center}
\begin{tabular}{cc}
\includegraphics[width=5.7cm,angle=-90]{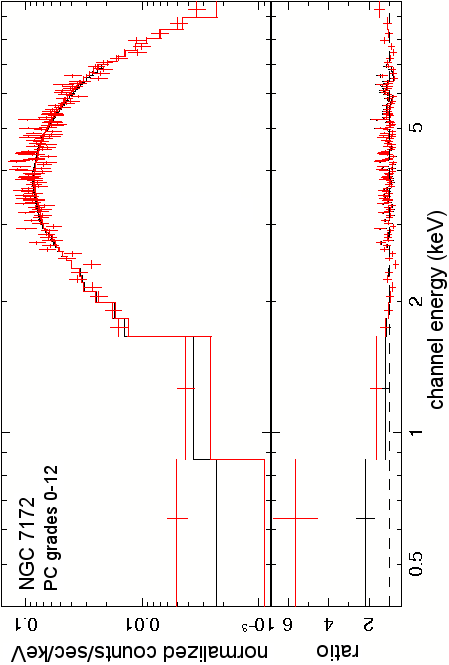} & 
\includegraphics[width=5.7cm,angle=-90]{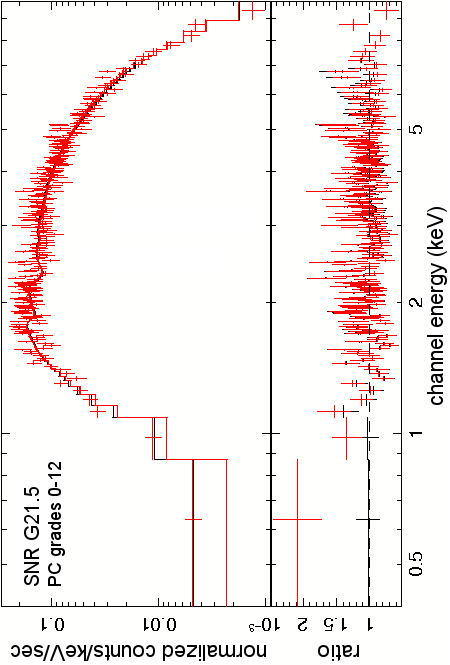}\\
\includegraphics[width=5.7cm,angle=-90]{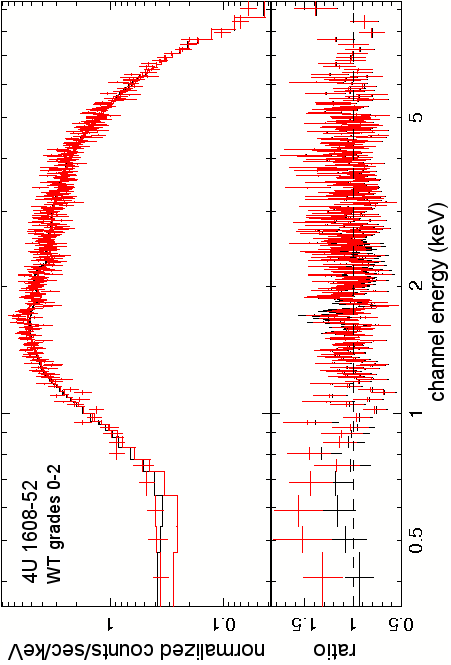} &
\includegraphics[width=5.7cm,angle=-90]{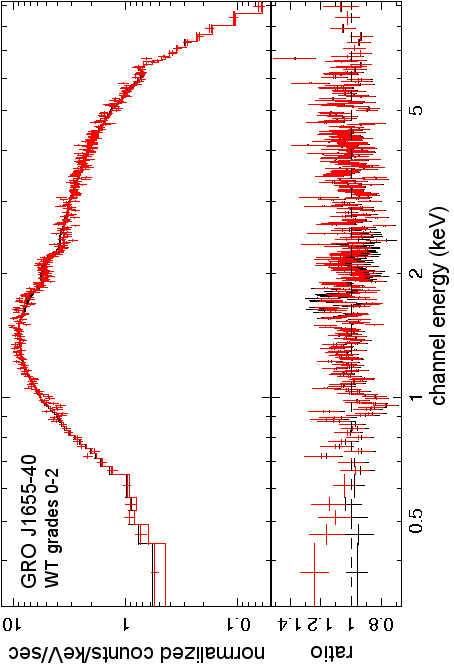}\\
\end{tabular}
\end{center}
\caption[]{Best fits of absorbed sources ($N_H > 5\times 10^{21}$ cm$^{-2}$)
in PC and WT modes showing the improvement in the modelling of the shelf with
the v009 RMFs. Top panel: (left) PC grade 0-12 spectrum of NGC 7172 using the
v009 (black) and previous v008 (red) response files. The model was a
{\scriptsize WABS*POWERLAW} model; (right) the same for the PC grade 0-12
spectrum of the SNR G21.5 using a {\scriptsize TBABS*POWERLAW}. The spectrum
was extracted using a 40'' circle region centred on the core of the
remnant. Bottom panel: (left) WT grade 0-2 spectrum of the X-ray binary 4U
1608-52 using the v009 (black) and previous v008 (red) response files. The
model was a {\scriptsize WABS*(POWERLAW+DISKBB)} model; (right) the same for
the WT grade 0-2 spectrum of the microquasar GRO J1655-40 in a low/hard state
using a {\scriptsize WABS*(POWERLAW+GAUSS)} model. }
\label{fig_shelf}
\end{figure*}

\subsubsection{The modelling of the shelf from photons above $\sim 2$ keV}
\label{shelf}

Before the release of the v009 RMFs, the spectral fits of heavily absorbed
sources (with $N_H$ typically larger than $10^{22}$ cm$^{-2}$) in either PC or
WT mode showed an underestimation of the modelled redistributed counts at low
energies corresponding to the shelf (see the red curves in
Fig.~\ref{fig_shelf}).  As discussed in Section 2.1.3, the physical origin of
the shelf, and hence its modelling, is complex.  To improve the CCD response
model, we decided to rescale the shelf for incident photons above 2 keV, since
this is a straightforward approach. The result of this rescaling significantly
improves the residuals at low energy when fitting spectra of heavily absorbed
sources in both modes (see the black curves in Fig.~\ref{fig_shelf}). While
the rescaling of the PC shelf did not change the QE at high energy because the
shelf is at least two orders of magnitude below the photo-peak, the rescaling
of the shelf in WT mode increased the QE by $\sim 10\%$ at 6 keV. This
increase of the high-energy QE in WT mode also gives a better agreement with
expectation of the grade ratios between the PC and WT modes (see
Section~\ref{QE}).  This new modelling was included in the release of v009
RMFs.

\begin{figure}[h]
\begin{center}
\begin{tabular}{c}
\includegraphics[width=5.8cm,angle=-90]{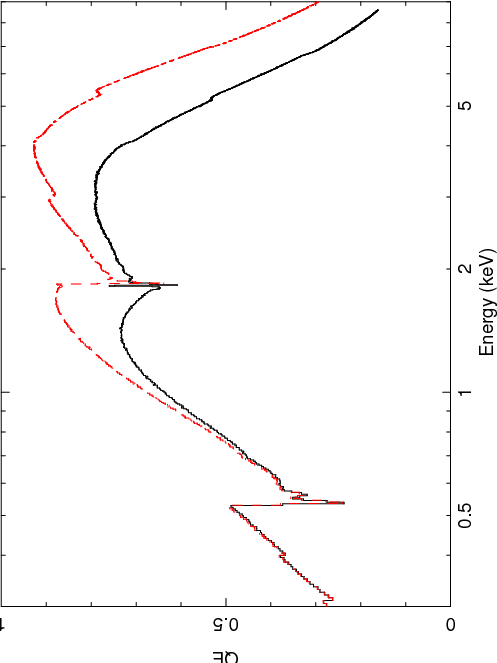}\\
\end{tabular}
\end{center}
\caption[]{Plot of the PC quantum efficiency using v007 RMF for: grade 0-12
(red) and grade 0 (black).}
\label{fig_QE_v007}
\end{figure}

\subsubsection{Origin of the low-energy shoulder from high-energy photons}
\label{shoulder}

For low-energy photons (below about 2 keV), the low-energy shoulder of the
line profile can be modelled successfully by charge losses at the interface
between silicon and oxide layer. However, the same process cannot completely
explain the low-energy shoulder seen in the line profile of high-energy
photons. Hence, the pre-launch PC and WT RMFs (v006) were not able fully to
model the line profile for high-energy photons (see the bottom panel in
Fig.~\ref{fig_low-E_ground}).

Tests performed after launch showed that an artificial increase of the split
threshold in the CCD response model results in an increase of sub-threshold
losses, and hence a better modelling of the shoulder. This empirical technique
was implemented in the v007 RMFs.  Even if the global result was to improve
significantly the residuals around the shoulder (see Fig.\,6a in Osborne et
al. 2005), this technique was not entirely satisfactory since residuals were
still present and the physical origin was unidentified.  The other drawback
was that the QE curve in PC mode showed discontinuities (smaller than $5\%$)
due to the use of different values of the threshold depending on the energy
range (see Fig.~\ref{fig_QE_v007}), although these small discontinuities had
no noticeable impact on the spectral fitting.

\begin{figure}[h]
\begin{center}
\begin{tabular}{c}
\includegraphics[width=8.5cm]{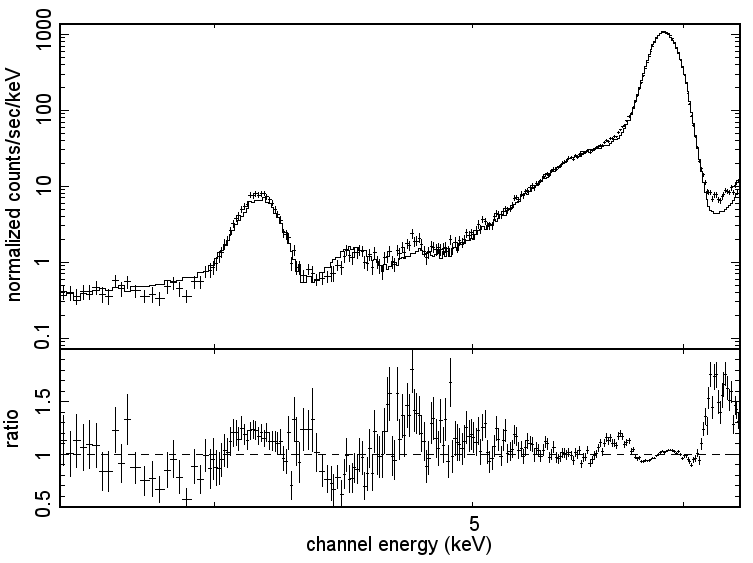}\\
\end{tabular}
\end{center}
\caption[]{PC grade 0-12 spectrum of the $^{55}$Fe source on the XRT camera
door fitted using the v009 RMF. The low energy shoulders of the lines are now
correctly modelled using the Pavlov \& Nousek formalism (compare to
Fig.~\ref{fig_low-E_ground} lower panel).}
\label{fig_shoulder}
\end{figure}

Pavlov \& Nousek (1999) have stressed that when a photon interacts in the
field-free region, the resulting charge cloud no longer has the profile of a
2-D Gaussian, because it experiences free diffusion in this region before
being distorted when penetrating the depletion region where the charge is
collected. The authors gave an analytical formalism to describe the resulting
shape of the charge cloud as a function of the depth of interaction.  The
resulting profile shows more extended wings which naturally favour the
increase of sub-threshold losses for a given threshold value.  The
implementation of the Pavlov \& Nousek formalism in our CCD response model
gives good results as shown in Fig.~\ref{fig_shoulder}.  This new modelling
was included in the release of v009 RMFs.

\subsubsection{Feature around 0.9-1.0 keV in WT mode}
\label{RMFv011}

\begin{figure}[h]
\begin{center}
\begin{tabular}{c}
\includegraphics[width=5.7cm,angle=-90]{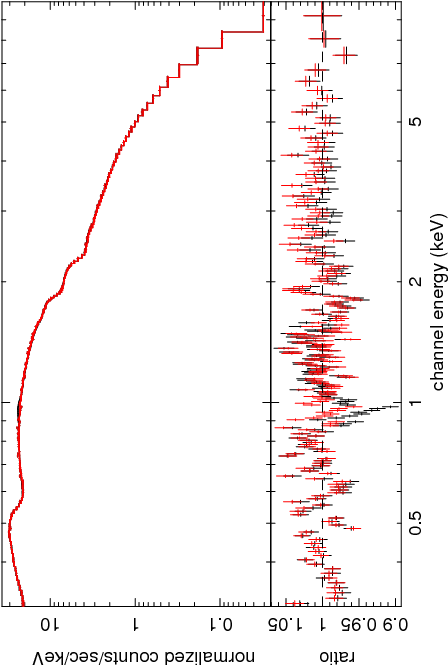}\\
\end{tabular}
\end{center}
\caption[]{WT grade 0-2 spectrum of Mkn\,421 from June 2006 data fitted using
  the v010 (black) and v011 (red) RMFs. The residuals around 0.9-1.0 keV
  (black points) are due to a RMF redistribution issue in v007-v010 WT RMFs.
  The use of the new v011 WT RMFs including a slight change in the loss
  function between 1 and 2 keV gives flatter residuals in this region (red
  points).  The spectral model is an absorbed bending power-law with $N_H =
  1.6\times 10^{20}$ cm$^{-2}$.}
\label{fig_Mkn421_feature}
\end{figure}

\begin{figure}[h]
\begin{center}
\begin{tabular}{c}
\includegraphics[width=6.cm,angle=-90]{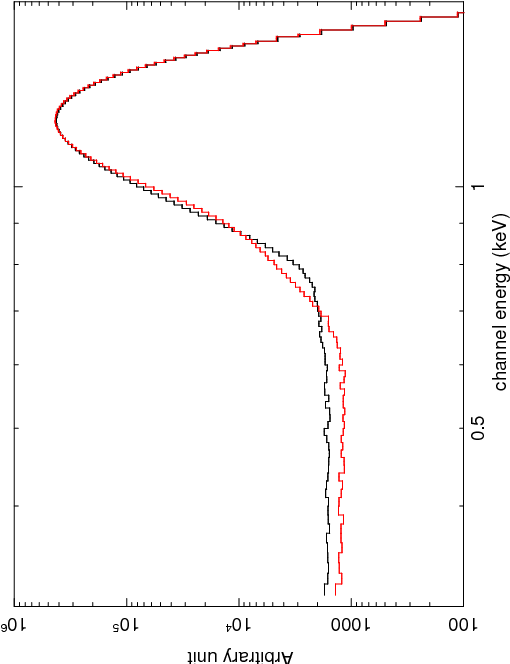}\\
\end{tabular}
\end{center}
\caption[]{Gaussian models with a value of $\sigma=0.1$ keV and an energy
  centroid of 1.2 keV folded through the RMF kernel: (black) CALDB WT grade 0
  RMF v010; (red) New WT grade 0 RMF v011. Note the change in the line profile
  due to a minor tweak of the loss function between 1 keV and 2 keV in order
  to remove the systematics observed around 0.9-1.0 keV.}
\label{fig_profile_0.9}
\end{figure}

Fits of high statistical quality WT spectra have revealed 10\% systematic
residuals around 0.9-1.0 keV, as shown in Fig.~\ref{fig_Mkn421_feature}.  An
ad-hoc dip was added to the WT v008-v010 ARFs around 0.9 keV in order to
suppress these residuals. However, this approach was not entirely
satisfactory. These systematic residuals are due to a RMF redistribution
issue. Indeed, as explained in Section~\ref{phys}, the surface charge losses
are modelled by a loss function $f$ so that its parameters are different in
different energy segments. 1\,keV corresponds to a boundary between two energy
segments for which there is a slight discontinuity in the description of the
loss function. We modified the parameters of the loss function in the 1-2 keV
range in order slightly to change the monochromatic line input profile, as
shown in Fig.~\ref{fig_profile_0.9} for 1.2 keV photons. New WT grade 0 and
0-2 RMFs (\emph{i.e.} v011) were computed using this new description. These
v011 RMFs give very good performance as shown by the red residuals in
Fig.~\ref{fig_Mkn421_feature} (also see Section~\ref{perfo}).

\subsection{Investigation of the QE shape from in-flight calibration}
\label{QE}
\subsubsection{Overview of the QE calibration}

The pre-launch QE was calibrated using measurements across the XRT energy
range (0.3-10 keV) made at the Leicester calibration facility. However, some
of these measurements were corrupted during the data taking process indicating
that some of the low energy QE values were not valid.

It is not straightforward to obtain a direct measure of the QE in orbit, since
the overall instrument spectral response depends on the mirror, filter
transmission and CCD responses. Early fitting results suggested that the
effective area needed to be rescaled between modes and grade selection due to
QE uncertainties. It was also necessary to introduce corrections around the
silicon and oxygen edges to obtain flat residuals when fitting spectra in both
PC and WT mode. As discussed in Section~\ref{offset}, the correction around
the oxygen edge was eventually found to be needed due to energy offsets in
both modes.

We discuss below possible explanations accounting for the changes implemented
in the ARFs and how they can be transfered to the CCD response model.

\subsubsection{The silicon K$\alpha$ edge and the low-energy QE shape}
\label{Siedge}

Since the detector consists of silicon bulk, and the effective area has a
maximum in the 1.5-2.2 keV energy band, it is essential to take special care
when modelling the Si K$\alpha$ edge (1.839 keV).

\begin{figure}[h]
\begin{center}
\begin{tabular}{c}
\includegraphics[width=5.7cm,angle=-90]{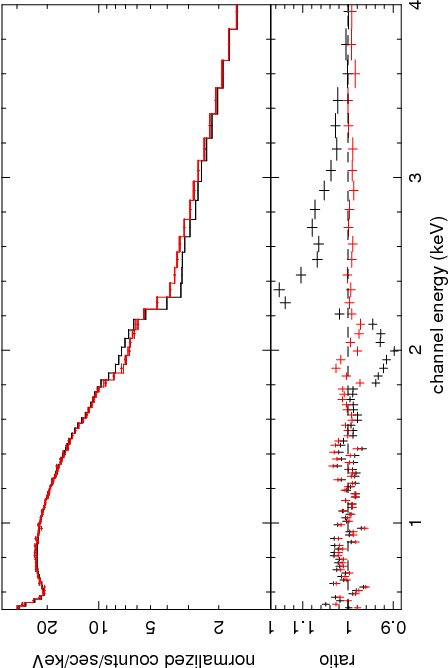}\\
\end{tabular}
\end{center}
\caption[]{Best fits of the WT grade 0 spectrum of Mkn\,421 using the WT grade
  0 v011 RMF together with the theoretical ARF (black) and the new ARF v011
  (red) which includes the correction of the CCD QE just above the Si
  K$\alpha$ edge (see text in Section~\ref{Siedge}). The residuals seen in
  black above 2.2 keV are due to the improper modelling of the Au M-shell
  (2.0-3.5 keV) edges in the theoretical ARF. A correction in the Au edge
  region (2.2-4 keV) was introduced in the ARFs after launch in order to
  flatten the residuals above 2.2 keV (red residuals).}
\label{fig_Si}
\end{figure}

Fits of several bright sources in both WT and PC modes using pre-launch RMFs
demonstrated that the modelling of the silicon edge was not completely
correct, as shown by the black residuals below 2 keV in Fig.~\ref{fig_Si} for
WT mode. The residuals above 2.2 keV are due to an improper modelling of the
Au M-shell edge region (2.0-3.5 keV) in the theoretical ARF. This ARF is made
of the filter transmission response {\scriptsize SWXFTRANS20010101V005.FITS}
and the effective area (v004), which is computed by a ray-tracing code
(Cusumano et al. 2005).

Initially, a correction was implemented in the v007-v010 PC and WT ARFs to
flatten the residuals in the 1.8-2.2 keV energy range. However, that
correction suffers from the effects of energy scale offsets which could not be
entirely corrected at those times.  Since then, we have been able to correct
the XRT data from these effects (see Section~3.2). This enabled us to
establish that the problem was related to an over-prediction of the QE in our
CCD response just above the Si K$\alpha$ edge (1.84-2.2 keV) rather than a
redistribution issue. A straightforward solution to correct the QE above the
Si edge was to implement a correction based on the XMM-MOS QE curve, which
shows a deeper profile at these energies. Unlike the {\it Swift}-XRT, the
XMM/MOS QE from the XMM-SAS 7.1.2 was calibrated with a synchrotron beam in
Saclay, allowing for accurate measurements around the Si edge. This
correction, along with a correction of the Au edge region (2.2-4 keV) in the
ARFs produced after launch, gives good results in WT mode (see the red curve
in Fig.~\ref{fig_Si}). A similar correction was implemented in the new v011 PC
ARFs.

Keay et al. (1995) stressed that the QE shape just above the Si edge was
strongly dependent on the electrode structure and composition of their JET-X
CCD.  They showed that the silicon contributes 65\% of the QE shape above the
edge, while the oxide and nitride of silicon contribute at 28\% and 7\%,
respectively.  As shown in Fig.~\ref{fig_pixel}, the description of the
electrode structure in the CCD pixel geometry used to generate the XRT RMFs is
relatively simple. In reality, there is an overlap of three different
electrodes.  An explanation for the QE over-prediction just above the silicon
edge could be that the thicknesses of the different materials in the CCD
geometry need to be more accurately specified. However, variation in thickness
of the Si and SiO$_2$ layers in the electrode and finger structures did not
allow us to converge towards adequate thickness values for these layers to
fully model the QE just above the silicon edge.

Nevertheless, from in-flight calibration we noticed that the total effective
area at low energy needs to be decreased in both WT and PC modes to obtain
correct flux levels and spectral parameters. Again, the electrode thicknesses
appeared to be under-specified, since more low-energy photons will be stopped
in the finger and electrode structures.

\begin{figure}[h]
\begin{center}
\begin{tabular}{c}
\includegraphics[width=6.cm]{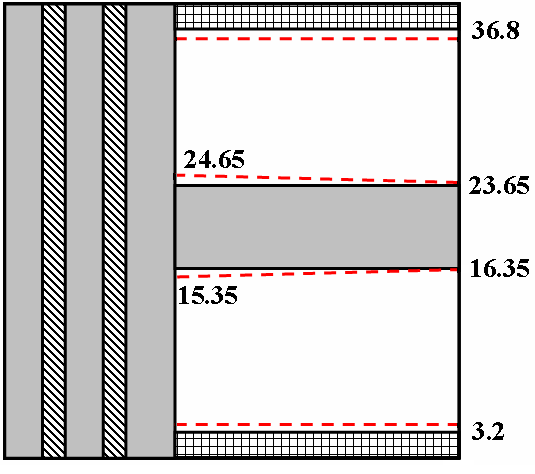}\\
\includegraphics[width=5.7cm,angle=-90]{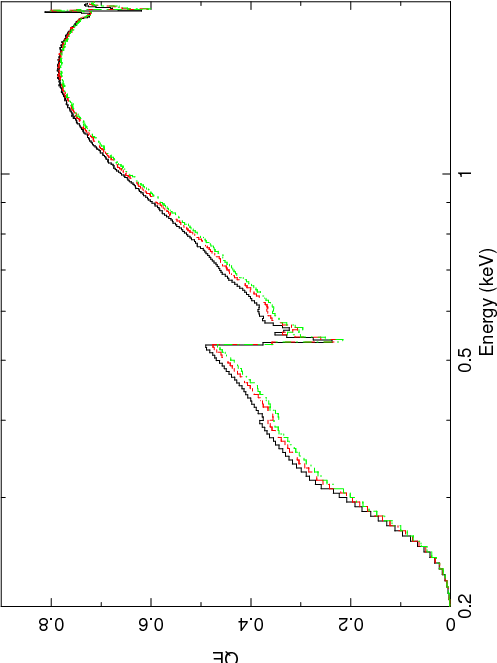}\\
\end{tabular}
\end{center}
\caption[]{(Top panel) Drawing showing the changes in the geometry of the
  finger and channel stop, corresponding to a reduction of the open area by
  $10.6\%$. (Bottom panel) Plot of the low-energy quantum efficiency for
  different configurations of the geometry of the CCD pixel: (black) using the
  default CCD geometry shown in Fig.~\ref{fig_pixel}; (red) with an increase
  of the width of each channel stop by 1$\mu$m; (green) with an increase of
  the width of each channel stop by 1$\mu$m and the finger structure as shown
  in the top panel.}
\label{fig_red}
\end{figure}

Another possible explanation accounting for the QE reduction at low-energy is
that the etching of the electrode and finger structure did not result in the
designed electrode widths, and some residual material was left as shown in
Fig.~\ref{fig_pixel}. The residual material acts as an extra absorption layer
for low-energy photons and so reduces the QE at low energy.  As a test we
reduced the open electrode area by slightly increasing the channel stop width
(by $1\,\mu$m) and the finger as shown in Fig.~\ref{fig_red} (top panel),
corresponding to an open area reduction of $10.6\%$.  The bottom panel in
Fig.~\ref{fig_red} shows a comparison of the WT grade 0-2 QE curve when using
the new geometries (in red and green) and that from the WT grade 0-2 RMF v010
(in black). A less than 10\% decrease of the QE is obtained at the energy of
the oxygen edge (0.543 keV). This is consistent with the ad-hoc reduction
implemented in the WT v007-v010 ARFs.

New WT ARFs (v011) were created by rescaling the theoretical ARF below 2 keV
using the green QE curve in the lower panel in Fig.~\ref{fig_red} and by
introducing the CCD QE correction just above the Si K$\alpha$ edge as
discussed above, while the Au M-shell region was corrected using celestial
continuum sources. These new WT grade 0 and 0-2 ARFs have similar shapes except
in the 1.5-1.84 keV energy range where the WT grade 0-2 ARF is rescaled down
by less than 2\% with respect to the WT grade 0 ARF in order to flatten
residuals below 2\% (see Fig. 3 in Godet et al. 2008).
Figs.~\ref{fig_Si} (red curve) \& \ref{fig_Mkn421_offset} show WT spectra of
Mkn\,421 fitted using an absorbed bending power-law and the new v011 response
files. This plot shows the very good performance of the v011 WT RMFs/ARFs,
since the residuals across the 0.3-10 keV energy range are better than 3\% and
the systematic errors are less than 2\% (also see Section~\ref{perfo}).

\subsubsection{Discrepancy in the effective area between PC and WT modes}
\label{phys_issue}

\begin{figure}[h]
\begin{center}
\begin{tabular}{c}
\includegraphics[width=7.cm]{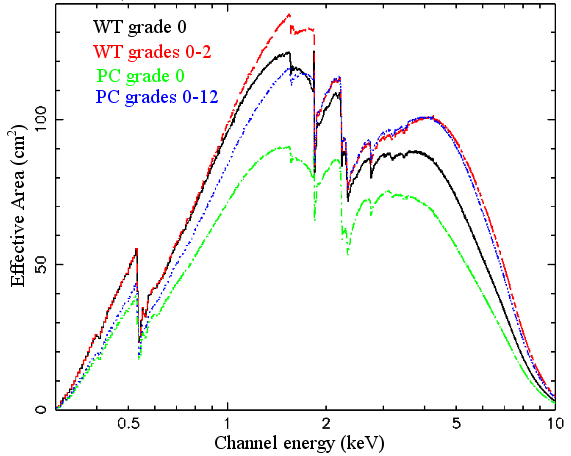} \\
\includegraphics[width=7.cm]{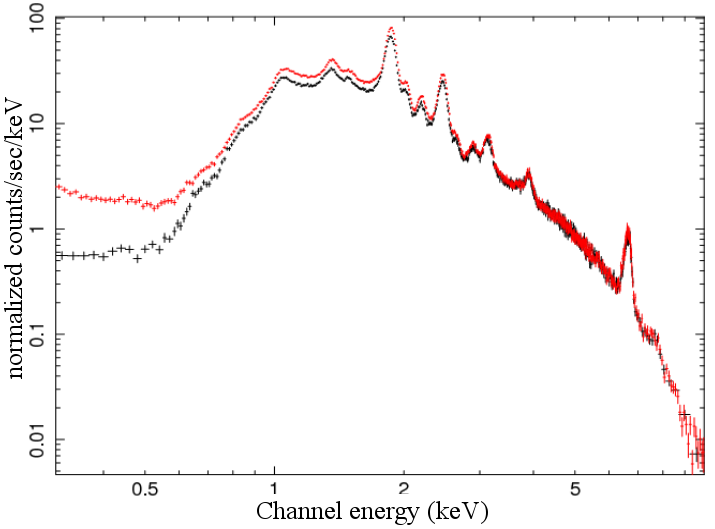} \\
\end{tabular}
\caption{(Top panel) Comparison of the XRT total effective area (EA) in PC and
  WT mode using the v011 response files. The PC grade 0 EA is very low when
  compared to the PC grade 0-12 and WT EA even at low energy ($E < 1$ keV)
  where we expect to have a similar QE level whatever the mode and grade
  selection. Below $\sim 1.84$ keV the PC grade 0-12 EA is less than the WT
  grade 0 EA, whereas we expect the WT grade 0 events to contain the PC grade
  0, 1 and 3 events.  (Bottom panel) PC (black: grades $0+1+3$) and WT (red:
  grade 0) spectrum of the whole remnant Cas A. The PC and WT data were
  processed using a central threshold of 80\,DN and a split threshold of
  80\,DN. For both spectra, we used the same size of the extraction region.
  There is a clear discrepancy between the two spectra between 1 and 3 keV
  which cannot be due to differences in the response redistribution between PC
  and WT mode.}
\label{fig_CasA_EA}
\end{center}
\end{figure}

\begin{figure*}
\begin{center}
\begin{tabular}{cc}
\includegraphics[width=6.5cm,angle=-90]{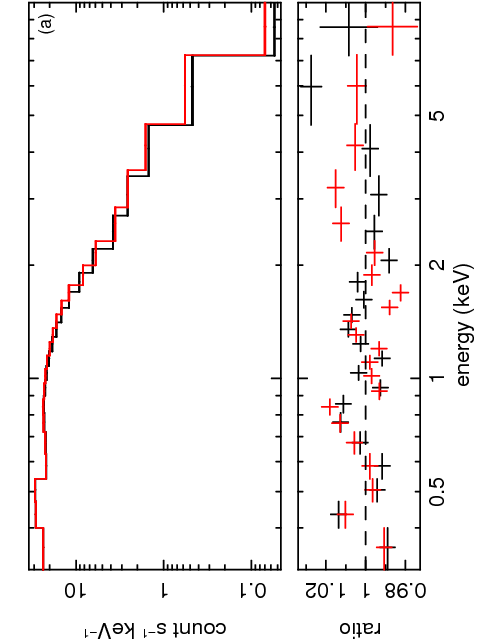} &
\includegraphics[width=6.5cm,angle=-90]{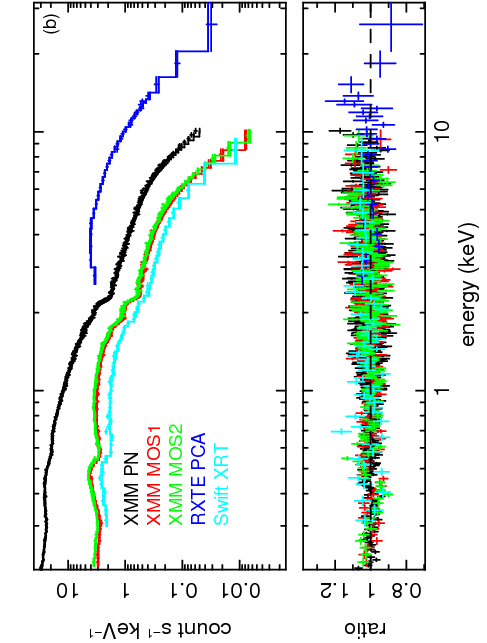}\\
\includegraphics[width=6.6cm,angle=-90]{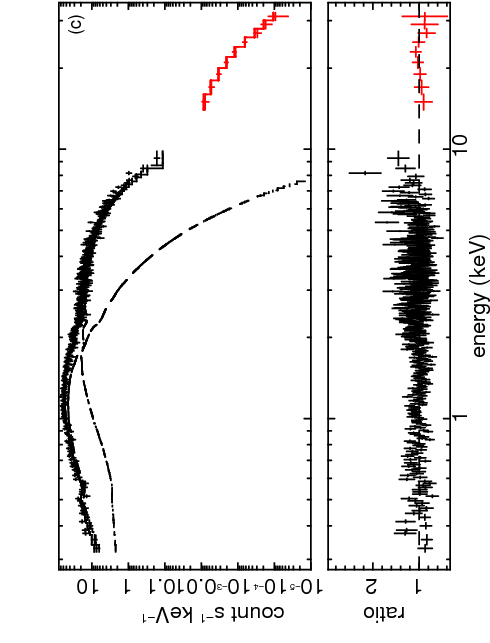} & \\
\includegraphics[width=6.7cm,angle=-90]{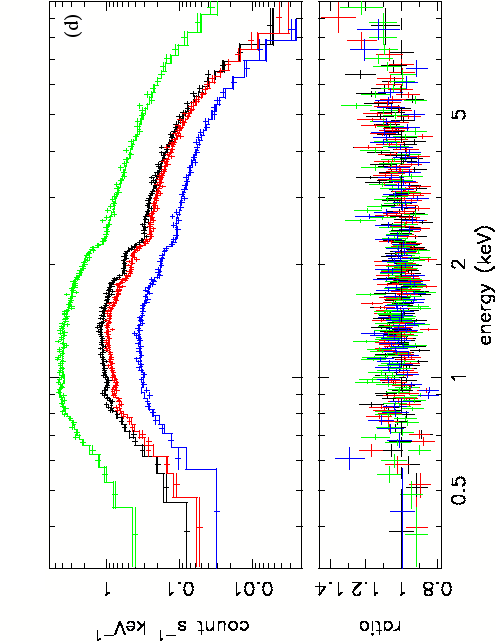} & 
\includegraphics[width=6.7cm,angle=-90]{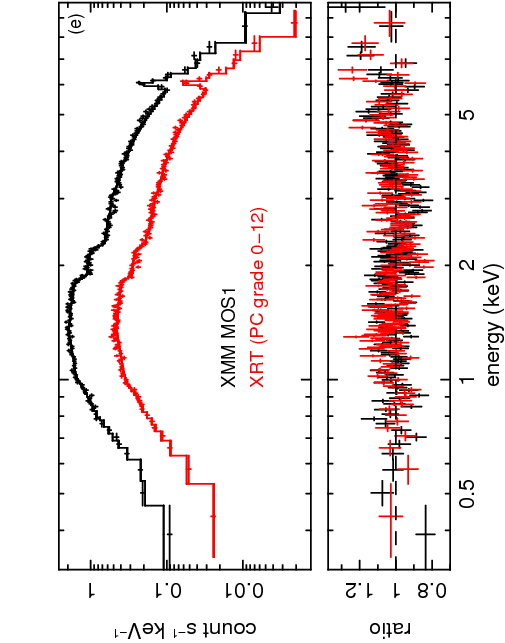}\\
\end{tabular}
\caption{Plots showing the best fits to different celestial sources when using
  the v011 PC and WT response files. (a) WT grade 0 (black) and grade 0-2
  (red) spectra of the blazar Mkn421 observed in June 2006. The spectral model
  is an absorbed bending power-law (i.e. {\scriptsize WABS} *
  exp[-$\alpha$+$\beta$ ln(E)]) with the column density fixed at
  $N_H=1.6\times 10^{20}$ cm$^{-2}$ (Kalberla et al. 2005). Each spectrum
  contains over $1.6\times 10^6$ counts. The spectra were heavily rebinned to
  better emphase the statistical residuals obtained for such high statistical
  quality data. The residuals are better than 3\% over the XRT energy range.
  (b) Simultaneous {\it XMM-Newton}, {\it Swift} and {\it RXTE} observations
  of 3\,C273 performed in July 2005 showing the PN (black), MOS1 (red), MOS2
  (green), {\it RXTE} (blue) and XRT WT grade 0-2 (cyan) fit using a
  {\scriptsize WABS}*{\scriptsize POWERLAW + 2 zBBODY} model. (c) Simultaneous
  {\it Swift}-XRT and {\it Swift}-BAT observation of Sco X-1 performed in
  April 2008 showing the XRT WT grade 0-2 (black) and BAT (red) fit. The data
  were fitted using a {\scriptsize CONST*WABS} ({\scriptsize BBODYRAD +
  COMPTT}) model with {\scriptsize CONST} fixed to 1 for the XRT data and left
  free when fitting the BAT data ({\scriptsize CONST} = $0.93\pm 0.04$, the
  errors being given at 2.71\,$\sigma$). The XRT/WT spectrum was extracted
  using an annulus region with an inner radius of 25 pixels since the data
  were heavily piled-up due to the high count rate of the source (2020 counts
  s$^{-1}$ when using a full circle extraction region). (d) Fits of PSR
  0540-69 (the pulsar and the nebula): (blue) {\it Swift}-XRT PC grade 0-12;
  (green) XMM-PN pattern 0-4; (red) XMM-MOS2 pattern 0-12; (black) XMM-MOS1
  pattern 0-12. The model is a {\scriptsize WABS}({\scriptsize VNEI +
  POWERLAW}) model. (e) Fits of the cluster PKS 0745-19: (red) {\it Swift}-XRT
  PC grade 0-12; (black) XMM-MOS1 pattern 0-12. The model is a {\scriptsize
  WABS}*{\scriptsize MEKAL} model with the abundances left free.  }
\label{fig_Mkn421_offset}
\end{center}
\end{figure*}

We expect that differences in the effective area between PC and WT modes would
be due to the differences in the event grading in these two modes.  However,
it was necessary to scale down the effective area in PC mode (for both grade 0
and 0-12) by a factor larger than expected from the simple differences in the
event grading, in order to reproduce the correct flux levels for the same
calibration targets (see the top panel in Fig.~\ref{fig_CasA_EA}; see Section
3.2 in Godet et al. 2008). Indeed, while the shape of the PC grade 0-12
effective area is similar to that of the WT grade 0-2 effective area above the
Si K$\alpha$ edge (as expected from the event grading), the WT grade 0
effective area is larger than the PC grade 0-12 effective area below the Si
K$\alpha$ edge. The event grading in PC and WT mode implies the opposite,
since WT grade 0 events only contain PC grade 0, 1 \& 3 events. The reason why
the PC grade 0-12 effective area is less than the WT grade 0 effective area
below $\sim 1.84$ keV remains unclear.

There is also a difference in the effective area at low energy (below 0.5 keV)
between the PC grade 0 and WT grade 0 effective areas.  The fact that the WT
grade 0 events contain the PC grade 0, 1 \& 3 events may account for the
discrepancy. However, the lower panel in Fig.~\ref{fig_CasA_EA} shows that
there are some differences in the Cas A spectra between WT grade 0 and PC
grades $0+ 1 + 3$ in the 1-3 keV band. To extract these spectra, we used 2005
data collected prior to the apparition of the bad columns, and the data were
processed using the same central (80 DN) and split (80 DN) event
thresholds. Below $\sim 1$ keV, the differences observed between the two
spectra may be partially due to differences in the kernel redistribution
between the modes (see Fig.~\ref{fig_spec_WT}). That suggests that the
observed differences in effective area at low energy cannot be due to simple
differences in the event grading.

%################################
\section{In flight spectroscopic performance and caveats}

We discuss in this section the in-flight spectroscopic performance of the v011
response files when fitting data collected at a substrate voltage of $V_{\it
ss}=0$\,V. For all the fits shown below, the PSF and exposure map corrections
were taken into account in the ARFs using the version 5.5 of the task
{\scriptsize XRTMKARF} of the {\scriptsize XRTDAS} software package (Capalbi
et al. 2005). We compare the fitting results with those found using other
in-orbit X-ray instruments. We also discuss current caveats in the XRT
spectral response: i) the line broadening due to the build-up of charge traps
on the CCD over time and methods to handle it; ii) the impact of the permanent
increase of the substrate voltage from 0\,V to 6\,V on the performance of the
v011 response files.

\begin{sidewaystable*}
\caption{Summary of the spectral parameters obtained when fitting {\it
Swift}-XRT, {\it XMM-Newton}, {\it Chandra} and {\it RXTE} data for several
continuum sources.}
\label{tab_intercal}
\begin{center}

\hspace{-4cm}\vspace{0cm}\begin{tabular}{|c|c|c|c|c|c|c|c c c|c|}
\hline
Mode & Grade & Source  & Model  & $N_H$ & $\Gamma$ or/and &
$F_{\mathrm{Obs}}$ [0.3-10 keV] &  & {\it XMM-Newton} results & & {\it RXTE} results\\

    & & & & ($\times 10^{22}$ cm$^{-2}$) & $kT$ (keV) & ($\times 10^{-11}$
erg cm$^{-2}$ s$^{-1}$) & PN & MOS1 & MOS2 & \\

\hline
\hline

PC & 0-12 & PKS 2155-304 & {\scriptsize WABS*POW} & 0.0124 fixed & $2.538 \pm
0.025$ &  &  $2.676 \pm 0.007$ & $2.590 \pm 0.010$ & $2.560 \pm 0.010$ & \\

& & & & & & $(14.72 - 15.16)$ & $(15.82 - 15.95)$ & $(14.17 -14.36)$ &
$(15.11 - 15.31)$ &  \\

PC & 0 & & & & $2.523 \pm 0.028$ & $(14.86 - 15.39)$ & & & &\\
\hline
%===========================================================================

PC & 0-12 & PKS 0745-19 & {\scriptsize TBABS*MEKAL}$^*$ & $0.54\pm 0.01$ &
$6.60\pm 0.22$ &  & & $6.55 \pm 0.19$ & &\\

& & & &  & $0.64\pm 0.07$ & & & $0.74 \pm 0.06$ & &\\

& & & &  &  &  $(4.76 - 4.86)$ & & $(4.44 - 4.53)$ & &\\

PC & 0 & & &  & $6.54\pm0.25$  &  & &  & &\\
&  & & &  & $0.72 \pm 0.08$ &  & &  & &\\
&  & & &  &  &  $(5.05 - 5.18)$ & &  & &\\

\hline
%===========================================================================

PC & 0-12 & SNR G21.5 & {\scriptsize TBABS*POW}$^{**}$ & $3.21\pm 0.17$ & &  &
$3.06\pm 0.05$ & $3.25 \pm 0.08$ & $3.23 \pm 0.08$ &\\

&  &  & &  & $1.782 \pm 0.072$ &  &
$1.763 \pm 0.021$ & $1.792 \pm 0.036$ & $1.839\pm 0.037$ &\\

&  &  & &  & & $4.53 \pm 0.12$$^\dagger$  &
$4.08 \pm 0.03$$^\dagger$ & $4.45\pm 0.05$$^\dagger$ & $4.59\pm 0.06$$^\dagger$ &\\

PC & 0 &  &  & $3.35\pm 0.19$ & &  & & & &\\
&  &  &  & & $1.834 \pm 0.086$ &  & & & &\\
&  &  &  & &  & $4.72\pm 0.15$$^\dagger$  & & & &\\

\hline
%===========================================================================
PC & 0-12 & RX J1856.4-3754 & Chandra$^{c}$ & & & $1.020 \pm
0.031$$^\ddagger$ & $0.944\pm0.005$  & $0.950\pm 0.010$ & $1.009\pm 0.010$ & \\

PC & 0 & & & & & $0.994 \pm 0.018$$^\ddagger$ & & & & \\

\hline
\end{tabular}
\end{center}
\end{sidewaystable*}

\begin{sidewaystable*}
\caption{The second part of Table~\ref{tab_intercal}.}
\begin{center}
\hspace{1cm}\vspace{0cm}\begin{tabular}{|c|c|c|c|c|c|c|c c c|c|}
\hline
Mode & Grade & Source  & Model  & $N_H$ & $\Gamma$ or/and &
$F_{\mathrm{Obs}}$ [0.3-10 keV] &  & {\it XMM-Newton} results & & {\it RXTE} results\\

    & & & & ($\times 10^{22}$ cm$^{-2}$) & $kT$ (keV) & ($\times 10^{-11}$
erg cm$^{-2}$ s$^{-1}$) & PN & MOS1 & MOS2 & \\

\hline
\hline
%===========================================================================
WT & 0-2 & 3C273 & {\scriptsize WABS(POW} & 0.0179 fixed & $1.57 \pm 0.04$ &  & $1.646 \pm 0.011$ & $1.500 \pm 0.028$ & $1.533 \pm 0.025$ & $1.62\pm 0.03$ \\

&  &  &{\scriptsize+2*zBBODY}) &  & $0.050\pm 0.030$ &  & $0.074 \pm 0.003$ & $0.089 \pm 0.004$ & $0.072 \pm 0.004$ &  \\

&  &  & &  & $0.136\pm 0.060$ &  & $0.186 \pm 0.011$ & $0.304 \pm 0.015$ &
$0.266 \pm 0.014$ &  \\

&  &  & &  & & $16.1\pm 0.3$  & $14.7 \pm 0.1$ & $15.3 \pm 0.1$ & $15.3 \pm 0.14$ &  \\
&  &  & &  & & $10.2\pm 0.2$$^\dagger$  & $8.5 \pm 0.1$$^\dagger$ & $9.5 \pm
0.1$$^\dagger$ & $9.3 \pm 0.1$$^\dagger$ &  $10.2\pm 0.1$$^\dagger$ \\

WT & 0 &  & &  & $1.56\pm 0.04$ &  & & & &  \\

&  &  & &  & $0.046\pm 0.030$ &  & & &  &  \\

&  &  & &  & $0.129\pm 0.090$ &  & & & &  \\

&  &  & &  & & $16.5^{+0.3}_{-0.4}$  & & & &  \\
&  &  & &  & & $10.2\pm 0.2$$^\dagger$  & & & &  \\

\hline
\end{tabular}
    \begin{list}{}{}
      \item $^*$ The abundance parameter was left as a free parameter.
      \item $^{**}$ The spectra were extracted using a 40 arcsec radius circle
      for both {\it Swift}-XRT \& {\it XMM-Newton} data. Extended source ARFs
      were specially created to fit the XRT spectra.
      \item $^\dagger$ The values of the observed flux are given in the 2-10
      keV energy range.
      \item $^{c}$ The model is {\scriptsize CONST*TBABS(BBODYRAD+BBODYRAD)}
    (see Beuermann et al. 2006). All the parameters are fixed
    ($N_H=1.1\times 10^{20}\,\rm cm^{-2}$, $kT_1 = 62.8$ eV and $kT_2 =
    32.3$ eV) except the constant factor. The lowest temperature
    black-body component has a minor impact in the XRT energy range. It
    was introduced by Beuermann et al. (2006) to fit the EUVE data as well
    as the Chandra data.
      \item $^\ddagger$ Values of the constant factor for the model described
      in Beuermann et al. (2006)
    \end{list}
\end{center}
\end{sidewaystable*}

\subsection{Spectroscopic performance}
\label{perfo}

Figure~\ref{fig_Mkn421_offset} (panel a) shows the best fit of the high
statistical quality WT spectra of the blazar Mkn\,421 observed in June 2006
using the v011 response files. Each spectrum contains more than $10^6$
counts. The residuals across the XRT energy range are below 3\%. The fits of
high statistical quality PC spectra show that residuals are less than 5\%
across the XRT energy range (see Godet et al. 2008).

The existence of several X-ray instruments (XMM-MOS, XMM-PN, {\it Chandra},
{\it RXTE}-PCA, {\it Suzaku}-XIS) covering an energy band similar to that of
the XRT enabled us to perform cross-calibration using different continuum and
line sources (see Fig.~\ref{fig_Mkn421_offset} panels b, d \& e and
Table~\ref{tab_intercal}; see Plucinsky et al. (2008) for a cross-calibration
work on the SNR E0102-723). Figure~\ref{fig_Mkn421_offset} shows that the XRT
spectral fit residuals are comparable with those of other instruments. The
spectral parameters derived using XRT data are consistent with those derived
using {\it XMM-Newton}, {\it Chandra} and {\it RXTE} data (see
Table~\ref{tab_intercal}). The XRT fluxes from PC grade 0 events seem to be
slightly higher up to 7\% when compared to those found from XRT PC grade 0-12
and other instruments, indicating that the PC grade 0 effective area may need
to be rescaled in the future. The WT grade 0 and grade 0-2 fluxes derived from
the fits of our WT calibration sources agree with each other within the error
bars (see Table~\ref{tab_intercal}).

The spectral analysis of joint BAT-XRT GRB spectra during overlapping time
intervals showed relatively good agreement between the two instruments, within
10\% in most cases. Figure~\ref{fig_Mkn421_offset} (panel c) shows the joint
fit of the BAT and XRT data of Sco X-1. The data were fitted using a
{\scriptsize CONST*WABS(BBODYRAD+COMPTT)} model. The constant factor
({\scriptsize CONST}) was fixed to 1 for the XRT data and left as a free
parameter for the BAT data ({\scriptsize CONST} = $0.93\pm0.04$, the errors
being given at 2.71\,$\sigma$). The residuals observed for the XRT spectrum in
Fig.~\ref{fig_Mkn421_offset} (panel c) above 8 keV correspond to Ni and Au
L-shell fluorescence.  Discrepancies larger than 10\% between the BAT and XRT
flux normalisation and/or large residuals at high-energy (above 5 keV) may
mean that: i) the spectral model used is not suitable to fit the data; ii) the
XRT data may be piled-up (pile-up is dependent on the spectral shape of the
source. However, a rough limit for pile-up to become an issue is 100 counts
s$^{-1}$ in WT mode and 0.6 counts s$^{-1}$ in PC mode for powerlaw-like
spectra); iii) instrumental lines such as Ni K$\alpha$ \& K$\beta$ and Au
L-shell fluorescence lines may be present above 8 keV (see Moretti et
al. 2008b).

Our current understanding of the XRT response at $V_{\it ss}=0$\,V implies a
systematic error of less than 3\% in both WT and and PC mode in the 0.3-10 keV
energy band and better than 10\% in absolute flux (see Godet et al. 2008).

\subsection{Line broadening}
\label{trap}

In Section~\ref{CTI}, we discussed the effect of the increase in CTI (due to
charge traps too shallow to be individually identified) which can cause an
energy scale change if not corrected. Besides this CTI, there is also a
build-up of deeper charge traps due to high-energy proton and radiation damage
to the CCD (the imaging area, the store frame area and the serial register),
which dominate the line broadening observed in XRT data (see
Fig.~\ref{fig_CTI}). The FWHM at 1.86 keV in the Cas A spectra has degraded
from 105\,eV in February 2005 to 131\,eV in June 2007. Over the same interval,
a 44\% width increase was measured at 5.9 keV from the on-board calibration
sources.  The most serious of these charge traps can cause a loss of up to 350
eV from the incident X-ray energy.

\begin{figure}[h]
\begin{center}
\begin{tabular}{c}
\includegraphics[width=5.5cm,angle=-90]{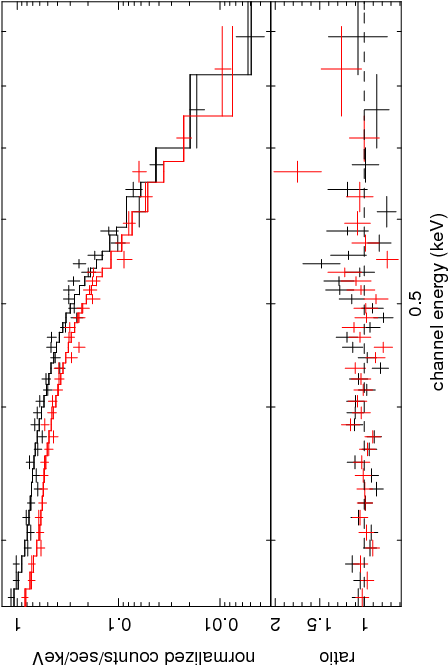}\\
\includegraphics[width=5.5cm,angle=-90]{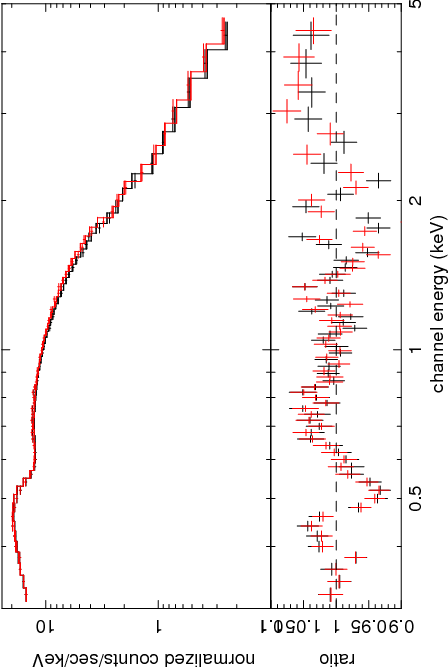}\\
\end{tabular}
\end{center}
\caption[]{Effects of the build-up of charge traps on the XRT spectral
  response for continuum sources. (Top panel) Evolution of the PC grade 0
  spectrum of the soft neutron star RX J1856.4-3754 over time: (black) data
  from February 2005 and (red) data from June 2007. The decrease in the flux
  normalisation between the two epochs by $\sim 22\%$ is due to the loss of
  events below the on-board central event threshold (see text in
  Section~\ref{trap}). (Bottom panel) WT grade 0 (black) and 0-2 (red) spectra
  of Mkn\,421 observed in March 2007 fitted using the v011 WT RMFs/ARFs. The
  spectra were fitted using an absorbed bending power-law as defined in
  Fig.~\ref{fig_Mkn421_offset}. The WT spectra contain more than $2\times
  10^5$ counts. The residuals around the instrumental edges are due to the
  energy shifting effect of charge traps (see text in Section~\ref{trap}). }
\label{fig_low2}
\end{figure}

Since launch, regular observations of the soft neutron star RX J1856.4-3754
have enabled us to monitor any significant low energy redistribution change in
the CCD response. The top panel in Fig.~\ref{fig_low2} shows the evolution of
PC grade 0 spectra of RX J1856.4-3754 between February 2005 and June 2007.
Both spectra were obtained using time intervals {\it when} the source was not
located on the bad columns.  Fitting the spectra with the model described in
Fig.~\ref{fig_low} gives a constant factor of $CONST=0.97\pm 0.03$ for
February 2005 data and $CONST=0.76\pm 0.02$ for June 2007 data. The use of a
simple {\scriptsize WABS*(BBODYRAD)} with the column density fixed at
$1.1\times 10^{20}$ cm$^{-2}$ gives a black-body temperature of $kT=62.8\pm
1.7$ eV for both datasets; which is consistent with the canonical value found
in the literature. A similar evolution is also observed in WT mode with a
constant factor of $CONST = 0.98\pm 0.03$ for February 2005 data and $CONST =
0.80\pm 0.03$ for June 2007 data.  All the errors quoted above are given at
$2.71\sigma$. Continuum spectra of Mkn\,421 from March 2007 WT data also
showed spectral fit residuals around the instrumental edges, especially the
oxygen edge (see the bottom panel in Fig.~\ref{fig_low2}), while no evidence
for such residuals was observed in June 2006 data.

\begin{figure}[h]
\begin{center}
\begin{tabular}{c}
\hspace{-0.8cm}\includegraphics[width=7.cm,height=5cm]{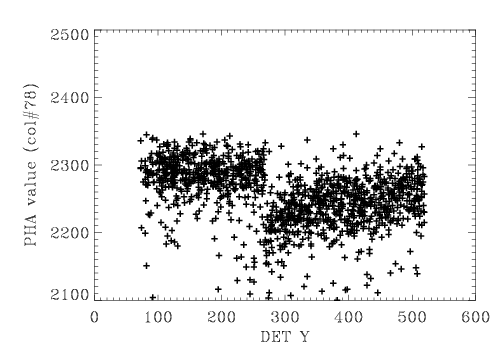}\\
\includegraphics[width=7.cm]{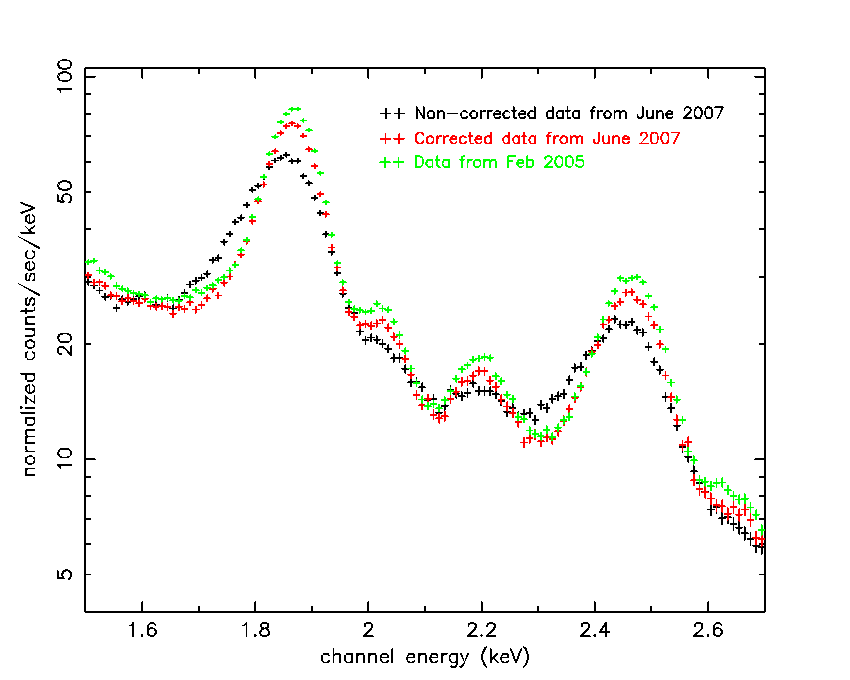}\\
\end{tabular}
\end{center}
\caption[]{(Top panel) PHA distribution of the Mn K$\alpha$ and K$\beta$ line
PC grade 0-12 events over the column DETX 78 of the CCD-22. The location of a
charge trap is clearly visible around row 271. Only the events above the trap,
i.e. above the row 271, are affected. (Bottom panel) WT grade 0-2 spectra of
the SNR Cas A: (black crosses) the spectrum from June 2007 data; (red crosses)
the same when the data were corrected from the effect of deep charge traps by
applying energy offsets in different columns; (green crosses) spectrum from
February 2005 data. The FWHM improves significantly after the correction of
the data.}
\label{fig_trap}
\end{figure}

All these changes are likely to be due to the build-up of charge traps over
time.  Charge traps are thought to be due to faults in the silicon crystalline
structure which trap a fraction of the charge passing through them during the
readout process. Therefore, only the events occurring above a trap in a given
column experience a charge loss.  The top panel in Fig.~\ref{fig_trap} gives
an example of a charge trap formed just after launch in the column DETX
78. The operation of the CCD at relatively high temperatures (from
$-$70$^\circ$C to $-$50$^\circ$C) may partially fill the shallowest traps
since the level of the thermally induced dark current is higher than that
expected at the designed operating temperature of $-100^\circ$C (see
Section~\ref{Vss}).

\begin{figure*}
\begin{center}
\begin{tabular}{cc}
\includegraphics[width=7.8cm,angle=0]{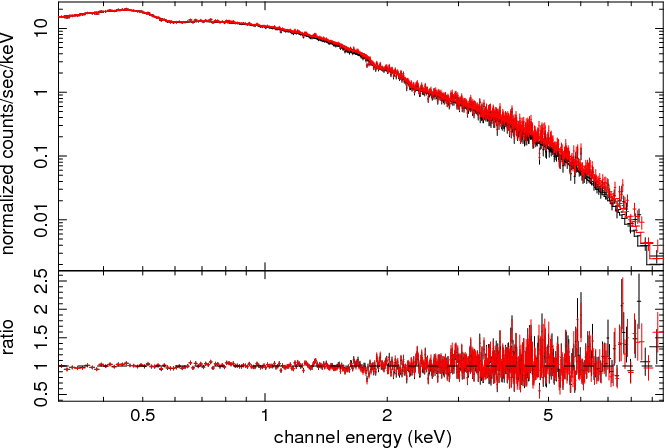} & 
\includegraphics[height=5.4cm,width=8.2cm,angle=0]{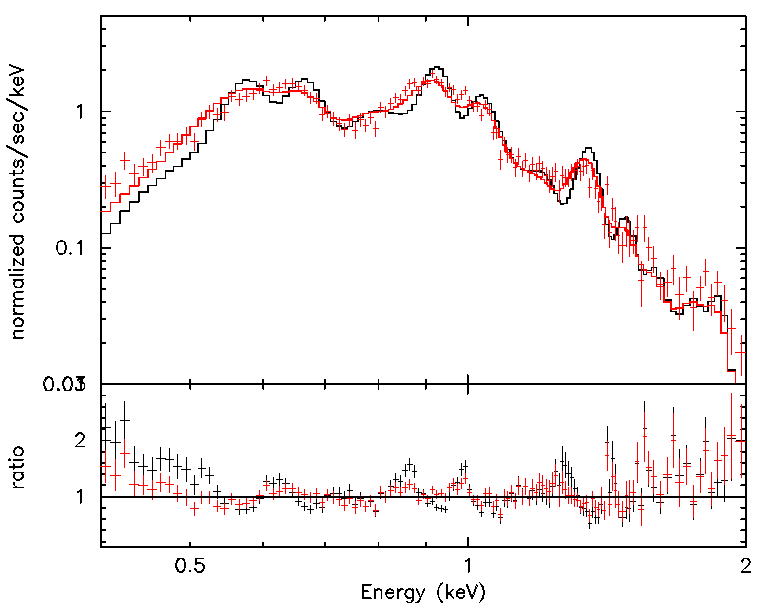}\\
\includegraphics[width=5.2cm,angle=-90]{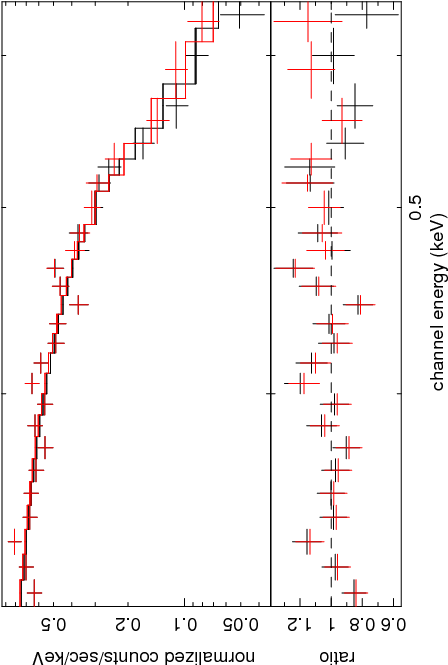} & \\
\end{tabular}
\end{center}
\caption[]{WT spectra of celestial sources using the v011 WT ARFs and/or the
  experimental broadened WT RMFs: (Top left panel) Fits of the Mkn\,421 data
  from March 2007 using the experimental broadened WT RMFs. The WT grade 0
  (black) and 0-2 (red) spectra contain more than $2\times 10^5$ counts and
  were fitted using the same model as given in Fig.~\ref{fig_low2} (bottom
  panel). (Top right panel) SNR 2E 0102-723 from June 2007 data (black: v011
  WT grade 0-2 RMF; red: new broadened WT grade 0-2 RMF). The WT grade 0-2
  spectrum contains more than $2.2\times 10^4$ counts. The spectral model is
  based on an XMM/RGS-PN model (see Plucinsky et al. 2008). The spectrum was
  extracted using data when the source was not located on the bad columns. We
  did not apply any PSF correction since the source is extended, and there is
  no tool in the ground software to generate extended ARFs.  (Bottom panel)
  RX\,J1856.4-3754 from June 2007. The spectra contain more than $2.9\times
  10^3$ counts. The spectral model is the same as given in
  Section~\ref{low_E}. The value of the constant factor is $C=0.90 \pm 0.02$
  for grade 0 (black) and $C=0.94\pm 0.03$ for grades 0-2 (red). All the data
  were processed with the new WT gain file {\scriptsize
  SWXWTGAINS0$_{-}$20010101v008.FITS} (see Section~3.2.4).}
\label{fig_other}
\end{figure*}

Trap-induced line broadening is difficult to model since it may depend on the
energy of the incident photons and the source intensity.  Another uncertainty
comes from the fact that the location and depth of the charge traps are
unknown. Because all XRT spectral response files released so far have an
incomplete modelling of the effect of CTI or traps, we urge caution in the
interpretation of apparently sharp spectral features observed in XRT data from
March 2007 onwards using the current calibration files.

Prospects to deal with the line broadening are under investigation:

\underline{\it Characterization of traps with the largest depth }- We are
experimenting with a technique based on the characterization of the location
and energy offset induced by charge traps with the largest depth. The bottom
panel in Fig.~\ref{fig_trap} illustrates how powerful this technique can be in
restoring the spectral resolution of the XRT; the Cas A Si line FWHM measured
from corrected WT data from June 2007 is 109 eV compared to 131 eV before
correcting the data (see Godet et al. 2007b).

\medskip

\underline{\it Broadening of the RMF kernel }- We are developing a new tool
which allows us to broaden the RMF kernel by convolving each of the 2400 RMF
spectra with a flux-conserving function $f$ with a CTI-dependent FWHM. All our
experiments so far focused on WT mode, but the principle should be similar in
PC mode. The line broadening is dominated by the effect of the deepest charge
traps, which shift a given spectrum to lower energy by some amount. It is also
important to take into account the fact that the amount of charge lost depends
on energy in a subtle and poorly understood way.

Since the {\it true} energy dependency is still unknown, we divided the 0.3-10
keV energy range into three ranges: 0.3-2 keV, 2-5 keV and 5-10 keV. Then, we
defined in each of the three ranges an ad-hoc shape of the function $f$;
\emph{i.e.} in each range we assumed that the shift induced by a trap is
constant and independent of the photon energy.  The function $f$ is defined as
the sum of two Gaussians, with their width, relative normalisation and
separation kept as free parameters to be optimised using data from celestial
sources, so that the FWHM of each RMF monochromatic line photo-peak of the
v011 WT RMFs, once convolved by $f$, is multiplied by a factor $B$ which
tracks the CTI increase over time.

To test this technique, we first calibrated the function $f$ using March 2007
data, and we computed experimental WT RMFs with a broadened kernel. As
illustrated in Fig.~\ref{fig_other} (the top left panel), which shows the best
fits of the March 2007 WT spectra of Mkn421 (with an average count rate of 45
count s$^{-1}$) using these RMFs along with the v011 WT ARFs, the residuals
around the instrumental edges are flattened when compared to
Fig.~\ref{fig_low2} (bottom panel). We also investigated whether variation of
the CCD temperature could have a significant impact on the trap properties,
and so the line broadening. To do so, we considered four temperature ranges
with a width of 5$^\circ$C from $-$70$^\circ$C to $-$52$^\circ$C using
Mkn\,421 data from March 2007 up to August 2007. We did not notice any
significant change in the residuals around the instrumental edges when fitting
the WT spectra using the experimental WT RMFs, even though the monitoring of
the on-board corner sources has revealed a trend with temperature (a $\sim
25\%$ increase in the FWHM at 5.9 keV from $-$70$^\circ$C to $-$50$^\circ$C).

\begin{figure}[h]
\begin{center}
\begin{tabular}{c}
\includegraphics[width=5.7cm,angle=-90]{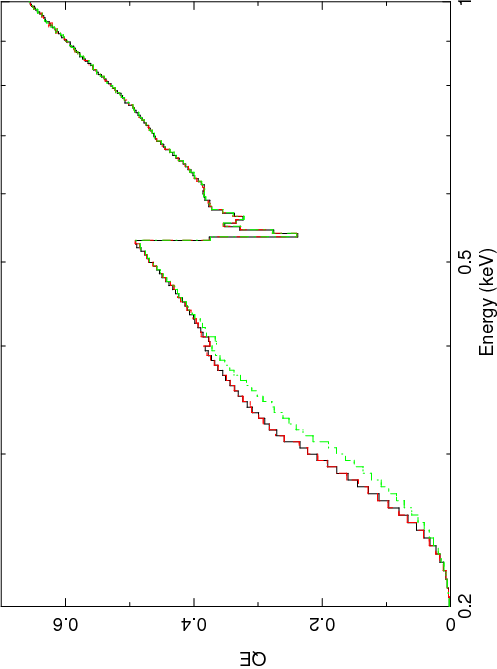}\\
\end{tabular}
\end{center}
\caption[]{WT grade 0-2 QE curves in the 0.2-1 keV energy band: (black) RMF
  v010, (red) RMF v011 and (green) broadened WT RMF v011. The loss of QE at
  low energy in the broadened WT RMF mimics the loss of events below the
  on-board central energy threshold due to the effect of charge traps.}
\label{fig_QE_broaden}
\end{figure}

We also tested these experimental WT RMFs on data from June 2007 up to
September 2007 for other celestial sources with lower count rates, such as the
SNR E0102-723 (1 count~s$^{-1}$) and the soft neutron star RX\,J1856.4-3754
(0.3 count~s$^{-1}$), to check the RMF kernel broadening and for any temporal
evolution of the line broadening. Indeed, the higher the source count rate,
the more likely the charge traps are to be filled; thus, it could be expected
that brighter sources show a less significant broadening. However, since the
XRT operates at much higher temperatures than it was designed for, most of the
small charge traps should be already filled. Figure~\ref{fig_other} shows the
very good performance of these experimental WT RMFs. Specifically we see no
intensity effect for sources between 0.3 count s$^{-1}$ to 70 count s$^{-1}$.
Moreover, their use enables us to retrieve in the case of RX\,J1856.4-3754
(June 2007 data) a normalisation consistent with that found using 2005-2006 PC
\& WT data (see Section~\ref{low_E}). This is because the new WT RMFs show a
slight decrease of the QE at low energy to take into account the fraction of
events lost below the on-board event threshold due to the effect of charge
traps (see Fig.~\ref{fig_QE_broaden}).

\subsection{Increase of the substrate voltage to 6V}
\label{Vss}

\begin{figure}[h]
\begin{center}
\begin{tabular}{c}
\includegraphics[width=5.8cm,angle=-90]{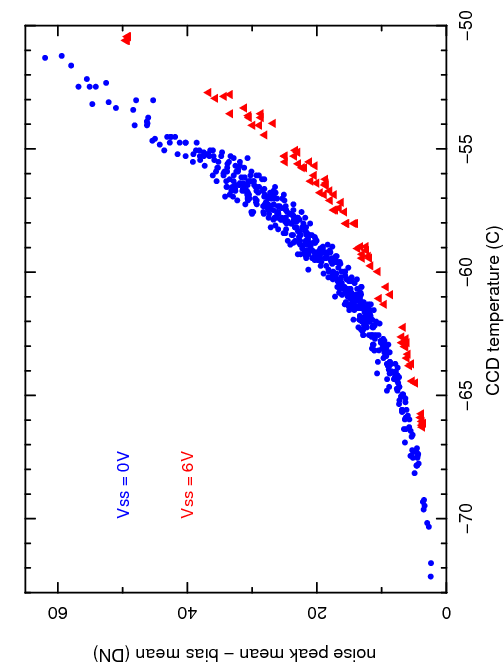}\\
\end{tabular}
\end{center}
\caption[]{Evolution of the thermally induced dark current as a function of
  the CCD temperature: (blue points) before the substrate voltage change to
  $V_{\it ss}=6$\,V; (red triangles) after the change.  The level dark current
  (given in DN) is computed as the difference between the mean of the noise
  peak and the mean of the bias level in the data. A clear reduction of the
  dark current can be seen due to the reduction of the volume of silicon in
  which carriers are generated is reduced.}
\label{fig_Vss}
\end{figure}

As discussed in Section~\ref{TEC}, the XRT operates at higher than expected
temperatures, resulting in significant thermally-induced noise appearing as
low energy events. From experiments performed at the Leicester calibration
facility on the {\it Swift} flight spare CCD-22 devices, we have demonstrated
that raising the substrate voltage to $V_{\it ss} = 6$\,V reduces the dark
current, since the volume of silicon in which carriers are generated is
reduced (Osborne et al. 2005, Godet et al. 2007b). Figure~\ref{fig_Vss}
clearly illustrates the benefit of raising the substrate voltage, since, for
the same level of dark current, it is now possible to operate the CCD at a
3-4$^\circ$C warmer temperature, and hence to collect useful science data up
to -50$^\circ$C, before excessive hot pixels compromise the data. The only
minor drawback of raising this voltage is that it induces a decrease of the
depletion depth, and therefore of the QE at high energy and just below the Si
edge ($\sim 1.5-1.84$ keV), due to the lower transparency of the silicon in
this energy range (see Fig. 4). The increase of the substrate voltage also
results in a change in the gain $C_0$, since the gain of the output FET is
also modified in this new configuration.

Numerical simulations performed using our CCD response model and laboratory
measurements made on a spare detector have shown that the QE decrease should
be small (less than 10\% at 6\,keV; Osborne et al. 2005, Godet et al. 2007b).
Short observations of two bright celestial targets with $V_{\it ss}$ set to
6\,V -- Cas A (2.2 ks in PC mode and 0.8 ks in WT mode) and the Crab (0.6\,ks
in WT mode) -- were also performed to estimate the on-board effective area
change and to create new PC and WT gain files. The QE changes were measured to
be $\sim 7\%$ at 6 keV and less than 10\% in the 1.5-1.84 keV range.  These
$V_{\it ss}=6$\,V gain files were released as version 007 prior to the
permanent change (see Table~\ref{tab_gain}).  The change in substrate voltage
has made it necessary to release two sets of gain files, now distinguished by
the characters `s0' and `s6' in their file names.

\begin{table}
\caption{Best fit parameters of the PC grade 0-12 spectra extracted using the
40 arc-second core of the SNR G21.5. }
\label{tab3}
\begin{center}
\begin{tabular}{|c|c|c|c|}
\hline
$V_{ss}$  & $N_H$  & $\Gamma$ & Normalisation$^\dagger$\\
(V) & ($\times 10^{22}$ cm$^{-2}$) & & ($\times 10^{-2}$ \\
    &                              & & ph keV$^{-1}$
cm$^{-2}$ s$^{-1}$) \\
\hline
0 & $3.26^{+0.17}_{-0.16}$ & $1.92\pm 0.07$ & $1.71^{+0.20}_{-0.18}$$^*$\\
6 & $3.38^{+0.14}_{-0.13}$ & $2.01\pm 0.06$ & $1.84^{+0.17}_{-0.15}$$^*$\\
\hline
\end{tabular}
    \begin{list}{}{}
      \item $^\dagger$ The fits were performed using a {\scriptsize
TBABS*POWERLAW} model with the abundance table given by Wilms et
al. (2000). No PSF correction was applied since the source is extended. 

\item $^*$ All the errors quoted above are given at $2.71\sigma$.
    \end{list}
\end{center}
\end{table}

\begin{figure}[h]
\begin{center}
\begin{tabular}{c}
\includegraphics[width=5.5cm,angle=-90]{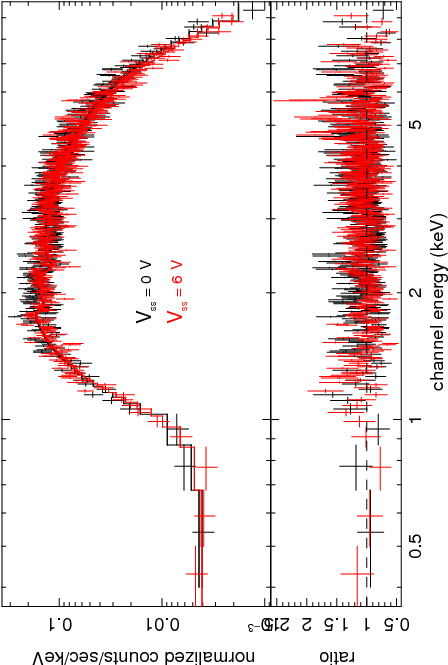}\\
\end{tabular}
\end{center}
\caption[]{Comparison of the residuals for PC grade 0-12 data of the SNR G21.5
  collected at $V_{\it ss}=0$ (black) and 6 V (red). The PC spectra were
  extracted using the 40 arc-second core of the remnant. The spectra were fit
  using the v011 PC RMFs and ARFs and the v008 PC gain files.  The PC spectra
  at $V_{\it ss}=0$ and 6\,V contain $9.4\times 10^3$ and $1.5\times 10^4$
  counts, respectively. The level of residuals is consistent between the two
  datasets as well as the spectral fitting parameters (see Table~\ref{tab3}).
  }
\label{fig_G21.5_Vss}
\end{figure}

Because the QE reduction was estimated to be rather small and the effect of
the operational change on the spectroscopic performance was demonstrated not
to be significant for most of the XRT observations (see
Fig.~\ref{fig_G21.5_Vss} and Table~\ref{tab3}), the substrate voltage was
permanently raised to 6\,V on-board on 2007 August 30.

Even though the QE change is small, and really noticeable only in WT spectra
with more than $10^5$ counts, the RMFs and ARFs need to be updated for both PC
and WT modes. An intense phase of re-calibration of the instrument is ongoing,
as indicated in Table\,1. The full details of the calibration of the $V_{\it
ss}=6$\,V response files will be addressed in a forthcoming paper.  By the
time this paper is published, new response matrices will be available that
take the QE change into effect properly.

%################################

\section{Conclusion}

We described in detail our Monte-Carlo simulation, computing the {\it
Swift}-XRT PC and WT RMFs. The response model is mainly based on a physical
description of the interaction of photons in the CCD. We showed how we used
in-flight calibration to improve the XRT spectral response (the low-energy
response, the line profile and the shelf) by implementing empirical
corrections when it was not possible to implement physical ones.  All the
changes allow us to describe the CCD response well. We showed that the v011
XRT response files, calibrated using data collected at $V_{\it ss}=0$ V, give
good performance on continuum and line sources in both PC and WT mode when
compared to other X-ray instruments in the 0.3-10 keV energy band (the
recommended bandpass) with a systematic error of less than 3\% in both modes
over 0.3-10 keV and better than 10\% in absolute flux.

We also showed that the XRT spectral response calibration was complicated by
various effects, such as energy offsets, related to the way the CCD is
operated in orbit. We described how these effects can be corrected in the
ground processing software thanks to the task {\scriptsize XRTWTCORR} in WT
mode and the task {\scriptsize XRTPCBIAS} in PC mode, and the release of
temperature-dependent gain files.  We stressed that the CCD started showing
evidence of degradation of its spectroscopic performance (line broadening,
small residuals around the instrumental edges and change in the low-energy
response) when fitting spectra of celestial sources from the middle of
2007. This is due to the build-up of charge traps in the CCD, produced by
high-energy proton and radiation damage. We are developing RMFs with a
broadened kernel, enabling us to handle the line broadening over time. The
preliminary results in WT mode are promising. Before the release of broadened
kernel RMFs in either mode, we recommend that users should be cautious in any
interpretation of the data.

In order to improve further the XRT spectral performance, the substrate
voltage was raised permanently to $V_{\it ss}=6$\,V on-board on 2007 August
30. This change resulted in a slight decrease of the QE at high energy ($E >
3$ keV) and just below the Si edge ($1.0-1.8$ keV). This QE change is really
only noticeable when fitting high statistical quality spectra with more than
$10^5$ counts. An intense phase of re-calibration of the instrument is
on-going to update the spectral response files. In the meantime, we recommend
the use of the v011 response files, and we advice the user to be cautious in
the spectral analysis of XRT data until the response files are updated.

\begin{acknowledgements}

OG, APB, JPO, AFA, KLP gratefully acknowledge STFC funding. This work is
supported at INAF by funding from ASI through grant I/011/07/0.

\end{acknowledgements}

%################################

\end{document}